\documentclass[PAPER, cernpreprint, texlive=2016, paper=a4, USenglish, texmf]{atlasdoc}
 
\usepackage{atlaspackage}
\usepackage{atlasbiblatex}
 
\usepackage{atlasphysics}
\usepackage{multirow}
 
\graphicspath{{logos/}{figures/}}
 
\usepackage{ANA-HDBS-2018-34-PAPER-defs}
\usepackage{PMGRefs-defs}
 
\AtlasTitle{Search for Higgs boson pair production in the
two bottom quarks plus two photons final state in
$pp$ collisions at $\sqrt{s}=13$~TeV with the ATLAS
detector}

\AtlasAbstract{
Searches are performed for nonresonant and resonant di-Higgs boson production in the $b\bar{b}\gamma\gamma$ final state. The data set used corresponds to an integrated luminosity of 139~fb$^{-1}$ of proton--proton collisions at a center-of-mass energy of 13~TeV recorded by the ATLAS detector at the CERN Large Hadron Collider. No excess above the expected background is found and upper limits on the di-Higgs boson production cross sections are set.
A 95\% confidence-level upper limit of 4.2 times the cross section predicted by the Standard Model is set on $pp \rightarrow HH$ nonresonant production, where the expected limit is 5.7 times the Standard Model predicted value.  The  expected constraints are obtained for a background hypothesis excluding $pp \rightarrow HH$ production.
The observed (expected) constraints on the Higgs boson trilinear coupling modifier  $\kappa_{\lambda}$ are determined to be $[-1.5, 6.7]$ $([-2.4, 7.7])$ at 95\% confidence level, where the  expected constraints on  $\kappa_{\lambda}$ are obtained excluding $pp \rightarrow HH$ production from the background hypothesis. For resonant production of a new hypothetical scalar particle $X$ ($X \rightarrow HH \rightarrow b\bar{b}\gamma\gamma$), limits on the cross section for $pp \to X \to HH$ are presented in the narrow-width approximation as a function of  $m_{X}$ in the range $251 \leq m_{X} \leq 1000$~GeV. The observed (expected) limits on the cross section for $pp \to X \to HH$ range from 640~fb to 44~fb  (391~fb to 46~fb) over the considered mass range.
}
 
\PreprintIdNumber{CERN-EP-2021-180}
\author{The ATLAS Collaboration}
 
\AtlasRefCode{HDBS-2018-34}

\AtlasJournalRef{Phys. Rev. D 106 (2022) 052001}
\AtlasDOI{10.1103/PhysRevD.106.052001}

\hypersetup{pdftitle={ATLAS document},pdfauthor={The ATLAS Collaboration}}
 
\begin{document}
 
\maketitle
 
\tableofcontents

% The next lines are included from the .//sections/introduction.tex input file
\section{Introduction}
\label{sec:intro} 
 
Since the discovery of the Higgs boson in 2012~\cite{HIGG-2012-27,CMS-HIG-12-028}, one of the priorities of the ATLAS and CMS Collaborations has been to better understand the properties of the Brout--Englert--Higgs mechanism~\cite{Englert:1964et,Higgs:1964ia,Higgs:1964pj,Guralnik:1964eu,Higgs:1966ev,Kibble:1967sv}. The Higgs boson self-coupling provides information about the structure of the Higgs potential. 
It is possible to directly probe the self-coupling of the Higgs boson by studying Higgs boson pair ($HH$) production. Furthermore, any deviation of
the Higgs boson pair production rate from the Standard Model (SM) prediction would point to new physics beyond the Standard Model (BSM) and may be within the sensitivity reach of the proton--proton ($pp$) collision data collected at $\sqrt{s} = 13~\TeV{}$ during  \RunTwo{} of the Large Hadron Collider (LHC)~\cite{Huang:2016cjm}.
 
\begin{figure}[tbh!]
\centering
\subfloat[Trilinear coupling]{
\includegraphics[width=0.43\textwidth]{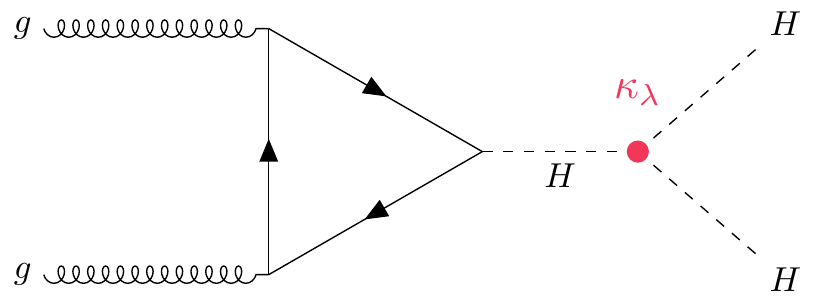} }
\quad\quad
\subfloat[Box diagram]{
\includegraphics[width=0.43\textwidth]{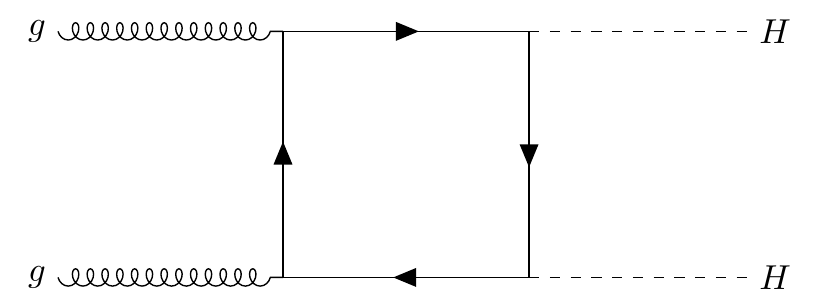} }
\caption[Feynman diagrams of gluon--gluon fusion di-Higgs production]{
The Feynman diagrams for the dominant gluon--gluon fusion production processes. In the Standard Model, the (a) trilinear coupling process, (b) the box diagram, and the destructive interference between the two processes,  contribute to the total cross section. In the figure, \klambda{} represents the Higgs boson trilinear coupling modifier. The quark content in the diagram is dominated by the top-quark contribution due to the large top-quark Yukawa coupling to the Higgs boson.
\label{fig:feyn_sm_hh} }
\end{figure}

At leading order (LO), the production of Higgs boson pairs via gluon--gluon fusion (\ggF) proceeds through the two diagrams shown in Figure~\ref{fig:feyn_sm_hh}. These diagrams interfere destructively, leading to a small production cross section~\cite{Dawson:1998py,Borowka:2016ehy,deFlorian:2013jea}. For 13~$\TeV$ $pp$ collisions and a Higgs boson mass $m_{H}=125.09~\GeV$~\cite{HIGG-2014-14}, the \ggF\ cross section, calculated at next-to-next-to-leading-order (NNLO) accuracy in the finite top-quark mass approximation (FTapprox), is
$\sigma_{HH}(\ggF) = 31.02^{+2.2\%}_{-5.0\%}\,\text{(Scale)}\,^{+4\%}_{-18\%}\,(m_{\text{top}})\,\pm 3.0\%\,(\alphas$+PDF)~fb~\cite{DiMicco:2019ngk,Shao:2013bz,deFlorian:2015moa,Baglio:2020wgt}, where ``Scale'' represents the uncertainty due to the finite order of the quantum chromodynamics (QCD) calculation, ``$m_{\textrm{top}}$'' the uncertainty related to the top-quark mass scheme~\cite{Baglio:2020wgt,Grazzini:2018bsd} which is added linearly to the Scale uncertainty, and ``$\alphas$+PDF'' the effect of uncertainties in the strong coupling constant and parton distribution functions.
 
\begin{figure}[tbh!]
\centering
\subfloat[$HHVV$ vertex]{
\includegraphics[height=0.2\textwidth]{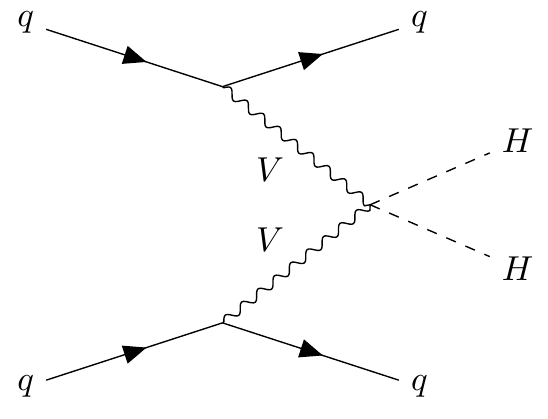} }
\quad
\subfloat[Trilinear coupling]{
\includegraphics[height=0.2\textwidth]{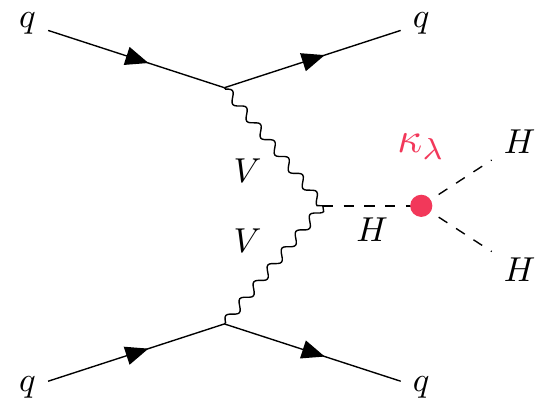} }
\quad
\subfloat[$VVH$ Production mode]{
\includegraphics[height=0.2\textwidth]{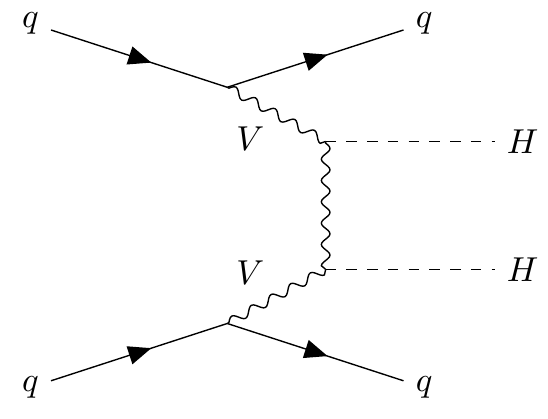} }
\caption[Feynman diagrams of Vector Boson Fusion di-Higgs production]{
The VBF production of Higgs boson pairs via (a) the $HHVV$ vertex,
(b) the trilinear coupling, and (c) the $VVH$ production mode. In the figure, \klambda{} denotes the Higgs boson trilinear coupling modifier.
\label{fig:feyn_hhvv} }
\end{figure}

The di-Higgs vector-boson fusion (VBF) production cross section, calculated at next-to-next-to-next-to-leading order (N3LO) for $m_{H}=125.09~\GeV$, is
$\sigma_{HH}(\textrm{VBF}) = 1.72^{+0.03\%}_{-0.04\%}\,\text{(Scale)}\,\pm 2.1\%\,(\alphas$+PDF)~fb~\cite{DiMicco:2019ngk}, which is one order of magnitude lower than the cross section of the \ggF\  process. The VBF production mode provides the analysis with additional sensitivity to the Higgs trilinear coupling
, as shown in Figure~\ref{fig:feyn_hhvv}. Both the \ggF\ and VBF production modes of Higgs boson pairs are considered as signal modes in this paper. The other production modes have lower cross sections~\cite{Frederix:2014hta} and are neglected.
 
Nonresonant enhancements to the Higgs boson
pair cross section can originate either from loop corrections involving new particles, such as light,
colored scalars~\cite{Kribs:2012kz}, or from non-SM couplings between the Higgs boson and other SM particles.
The nonresonant production cross section can also be altered by the trilinear self-coupling, $\lambda_{HHH}$, being different from the SM prediction, as discussed in Refs.~\cite{DiLuzio:2017tfn,Kribs:2017znd}.
Such an effect can be captured by a scale factor  $\kappa_{\lambda}$ defined as  $\kappa_{\lambda}= \lambda_{HHH}/\lambda^{\textrm{SM}}_{HHH}$, where $\lambda^{\textrm{SM}}_{HHH}$ is the SM value of the parameter.
 
\begin{figure}[tbh!]
\centering
\includegraphics[height=0.2\textwidth]{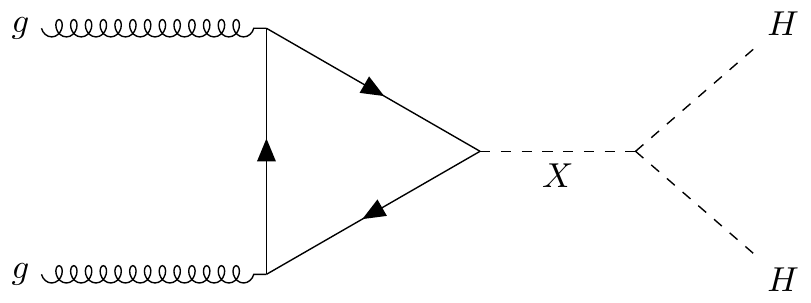}
\caption{Gluon--gluon fusion production of a heavy resonance decaying into a Higgs boson pair.}
\label{fig:feyn_resonance}
\end{figure}

In addition to the nonresonant enhancements, searching for resonant production
of Higgs boson pairs is well motivated.
Figure~\ref{fig:feyn_resonance} shows a \ggF{} production diagram possible in BSM theories predicting the existence of heavy scalar particles that can decay into a pair of Higgs bosons. Such theories
include models with two Higgs doublets~\cite{Lee:1973iz}, such as the minimal supersymmetric extension of the SM~\cite{Dimopoulos:1981zb}, twin Higgs models~\cite{Chacko:2005vw}
and composite Higgs models~\cite{Mrazek:2011iu},
adding a second complex scalar doublet to the Higgs sector.
Alternatively, the Randall--Sundrum
model of warped extra dimensions~\cite{Randall:1999ee} predicts spin-0 radions
that could couple to a Higgs boson pair.
 
This paper presents a search for di-Higgs production in the \yybb{} final state, including dedicated assessments of nonresonant and resonant contributions. The analysis considers the full \RunTwo data set of 139~$\ifb$ at 13~\TeV{}. For both the nonresonant and resonant $HH$ searches, the analysis employs a multivariate method designed to reject background processes, and the statistical results are obtained from a fit of the diphoton invariant mass, $\myy$. For the nonresonant search, data are divided into different categories based on the four-body invariant mass to target different $\kappa_\lambda$ ranges. The resonant search focuses on probing the existence of a narrow-width scalar particle $X$ in the mass range $251 < m_X < 1000$~\GeV{} decaying into a pair of Higgs bosons. The selection criteria depend on the mass of the probed scalar particle.  The main background processes are diphoton-plus-jets production and processes where a Higgs boson is produced and decays into a pair of photons.
In the context of the resonant search, nonresonant $HH$ production is considered as a background.
 
Previous results from the ATLAS Collaboration were obtained in this channel with an integrated luminosity of 36~$\ifb$ of data at 13~\TeV{} collected during Run 2, and they were found to be consistent with SM expectations within uncertainties~\cite{HIGG-2016-15}. The search for nonresonant enhancements of Higgs boson pair production set an observed (expected) 95\% confidence level (CL) upper limit on the $HH$ cross section of 0.73 (0.93)~pb, corresponding to 22 (28) times the SM prediction. The Higgs trilinear coupling was constrained to the range $-8.2 < \kappa_\lambda < 13.2$ at 95\% CL ($-8.3 < \kappa_\lambda < 13.2$ expected). A previous combination of searches for $HH$ pair production performed by  the ATLAS Collaboration with up to 36~$\ifb$ of 13~\TeV{} data provided constraints of $-5.0 < \kappa_\lambda < 12.0$ at 95\% CL~\cite{HDBS-2018-58}.
A search for enhancements due to the decay of a narrow-width scalar particle was performed in this channel as well, and results were given as a function of the resonance mass, $m_X$. The observed (expected) limits were between 1.1~pb and 0.12~pb over the range $260 < m_X < 1000$~\GeV.  A combination of searches  by  the ATLAS Collaboration with up to 36~$\ifb$ of 13~\TeV{} data for a narrow-width scalar resonance decaying into a $HH$ pair was performed and provided upper limits between 851~fb and 4.6~fb over the range $260 < m_X < 3000$~\GeV~\cite{HDBS-2018-58}.
The CMS Collaboration also set observed (expected) upper limits of 7.7 (5.2) times the SM prediction at 95\% CL on nonresonant enhancements of Higgs boson pair production in the \yybb{} final state with 137~$\ifb$ of 13~\TeV{} data~\cite{CMS-HIG-19-018}. In the same final state, the CMS Collaboration set 95\% CL upper limits on the production cross section of a narrow-width scalar particle between 4.2~pb and 0.23~pb over the range $260 < m_X < 900$~\GeV{} with 35.9~$\ifb$ of 13~\TeV{} data~\cite{CMS-HIG-17-008}.

% End of text imported from the .//sections/introduction.tex input file
 
% The next lines are included from the .//sections/atlas.tex input file
 
\section{The ATLAS detector}
 
\newcommand{\AtlasCoordFootnote}{
ATLAS uses a right-handed coordinate system with its origin at the nominal interaction point (IP)
in the center of the detector and the \(z\)-axis along the beam pipe.
The \(x\)-axis points from the IP to the center of the LHC ring,
and the \(y\)-axis points upwards.
Cylindrical coordinates \((r,\phi)\) are used in the transverse plane,
\(\phi\) being the azimuthal angle around the \(z\)-axis.
The pseudorapidity is defined in terms of the polar angle \(\theta\) as \(\eta = -\ln \tan(\theta/2)\).
The rapidity $y$ is defined in terms of the energy, the momentum and the polar angle \(\theta\):
$y=\frac{1}{2} \ln  \left( \frac{E+p \cdot \cos \theta}{E-p \cdot \cos \theta} \right)$.
The angular distance is measured in units of \(\Delta R \equiv \sqrt{(\Delta\eta)^{2} + (\Delta\phi)^{2}}\).}
 
The ATLAS detector~\cite{PERF-2007-01} at the LHC covers nearly the entire solid angle around the collision point.\footnote{\AtlasCoordFootnote}
It consists of an inner tracking detector surrounded by a thin superconducting solenoid, electromagnetic and hadron calorimeters,
and a muon spectrometer incorporating three large superconducting air-core toroidal magnets.
 
The inner-detector system (ID) is immersed in a \SI{2}{\tesla} axial magnetic field
and provides charged-particle tracking in the range \(|\eta| < 2.5\).
The high-granularity silicon pixel detector covers the vertex region and typically provides four measurements per track,
the first hit normally being in the insertable B-layer installed before Run~2~\cite{ATLAS-TDR-19,PIX-2018-001}.
It is followed by the silicon microstrip tracker, which usually provides eight measurements per track.
These silicon detectors are complemented by the transition radiation tracker (TRT),
which enables radially extended track reconstruction up to \(|\eta| = 2.0\).
The TRT also provides electron identification information.
 
The calorimeter system covers the pseudorapidity range \(|\eta| < 4.9\).
Within the region \(|\eta|< 3.2\), electromagnetic calorimetry is provided by barrel and
endcap high-granularity lead/liquid-argon (LAr) calorimeters,
with an additional thin LAr presampler covering \(|\eta| < 1.8\)
to correct for energy loss in material upstream of the calorimeters.
Hadron calorimetry is provided by the steel/scintillator-tile calorimeter,
segmented into three barrel structures within \(|\eta| < 1.7\), and two copper/LAr hadron endcap calorimeters.
The solid angle coverage is completed with forward copper/LAr and tungsten/LAr calorimeter modules
optimized for electromagnetic and hadronic energy measurements respectively.
 
The muon spectrometer (MS) comprises separate trigger and
high-precision tracking chambers measuring the deflection of muons in a magnetic field generated by the superconducting air-core toroidal magnets.
The field integral of the toroids ranges between \num{2.0} and \SI{6.0}{\tesla\metre}
across most of the detector.
A set of precision chambers covers the region \(|\eta| < 2.7\) with three layers of monitored drift tubes,
complemented by cathode-strip chambers in the forward region, where the background is highest.
The muon trigger system covers the range \(|\eta| < 2.4\) with resistive-plate chambers in the barrel and thin-gap chambers in the endcap regions.
 
Interesting events are selected by the first-level trigger system implemented in custom hardware,
followed by selections made by algorithms implemented in software in the high-level trigger~\cite{TRIG-2016-01}.
The first-level trigger accepts events from the \SI{40}{\MHz} bunch crossings at a rate below \SI{100}{\kHz},
which the high-level trigger reduces in order to record events at rate of about \SI{1}{\kHz}.\\
An extensive software suite~\cite{ATL-SOFT-PUB-2021-001} is used in the reconstruction and analysis of real
and simulated data, in detector operations, and in the trigger and data acquisition systems of the experiment.
% End of text imported from the .//sections/atlas.tex input file

% The next lines are included from the .//sections/DataSimSamples.tex input file
\section{Data and simulation samples}
\label{sec:simulation}
 
This analysis uses $pp$ collision data collected by the ATLAS experiment from 2015 to 2018 with proton beams colliding at a center-of-mass energy of $\sqrt s =13$~\TeV.
After data quality requirements~\cite{DAPR-2018-01} the full data set represents an integrated luminosity of $139.0 \pm 2.4$~fb$^{-1}$~\cite{LUCID2,ATLAS-CONF-2019-021}.
The mean number of inelastic $pp$ interactions per bunch crossing is 34.2~\cite{priv-SIMU-2020-01}.
 
Monte Carlo (MC) simulations are available for the signal as well as most background processes as detailed in the rest of this section.
The reducible backgrounds from final states with jets wrongly identified as photons ($\gamma$--jet and dijet backgrounds) are, however, estimated using a data-driven technique detailed in Section~\ref{sec:background_diphoton}.
 
Events from \ggF\ nonresonant $HH$ production were generated at next-to-leading-order accuracy in QCD with finite top-quark mass in both the real and virtual corrections (NLO FT)~\cite{Borowka:2016ehy}, using the \POWHEGBOX[v2]~\cite{Nason:2009ai} generator in the finite top-quark mass approximation~\cite{powheg-sm,powheg-kappa-lambda}  with the \PDFforLHC[15] parton distribution function (PDF) set~\cite{Butterworth:2015oua}.
The \PYTHIA[8.244]  generator was used for parton showering, hadronization and underlying-event simulation. \HERWIG[7.1.6] was used as an alternative generator to calculate the theory uncertainty from the parton shower.
Samples were generated for coupling modifier values $\kappa_\lambda = 1$ and $10$.
 
For \ggF\ nonresonant $HH$ production, a reweighting method based on the di-Higgs invariant mass $m_{HH}$ is used to provide predictions on the cross section at different \klambda{} values, starting from the existing $\kappa_\lambda = 1$ sample.
The reweighting method derives the scale factors as a function of $\kappa_\lambda$ in bins of $m_{HH}$ by performing a linear combination of samples generated at different $\kappa_\lambda$ values~\cite{ATL-PHYS-PUB-2019-007}. 
Histograms of the truth $m_{HH}$ distribution are produced for each $\kappa_{\lambda}$ sample and the distributions of the other relevant kinematic variables are obtained applying an event-per-event weight based on the ratio between the binned $m_{HH}$ distribution for the targeted $\kappa_{\lambda}$ by the binned $m_{HH}$ distribution for the SM ($\kappa_{\lambda} = 1$) sample.
 
This method was validated by comparing the event yields and the distributions of the relevant Higgs boson kinematic variables, including the $m_{\gamma \gamma}$ variable, of the sample generated with $\kappa_\lambda = 10$ to the sample generated with $\kappa_\lambda = 1$ and reweighted to $\kappa_\lambda = 10$. Good agreement is obtained in all categories. A systematic uncertainty in the range of 3\%--4\% is associated with the reweighting process, based on the maximum differences of signal yields observed in this validation. For each \klambda{} value, the inclusive cross section is normalized according to Ref.~\cite{Amoroso:2020lgh}. A fit with a second order polynomial to the MC prediction is performed in each analysis category, in order to parametrize the event yields as a function of \klambda{}.

For VBF nonresonant $HH$ production, \MGNLO[2.6.0]~\cite{Alwall:2014hca} was used to generate events at LO~\cite{Alwall:2014hca,vbf-pheno}.
The \NNPDF[3.0nlo] PDF set \cite{NNPDF} was used in the matrix element, interfaced to \PYTHIA[8.244].
The cross section of the VBF $HH$ process is evaluated at N3LO in QCD~\cite{Liu-Sheng:2014gxa,Dreyer:2018qbw,Dreyer:2018rfu}, as outlined in Section~\ref{sec:intro}.
Since samples were generated at LO for four values of the coupling modifier, $\kappa_\lambda = 0, 1, 2$ and $10$, the N3LO-to-LO cross-section ratio at the SM value is calculated and this factor is applied to the VBF $HH$ cross section.
These samples are used to derive a parameterization of the signal yields in the signal region as a function of $\kappa_\lambda$ by fitting a second-order polynomial to the MC predictions in each analysis category, as described in Section~\ref{subsec:BDTnon_resonant_selection}.
 
The production of a heavy spin-0 resonance $X$ via \ggF{} and its decay into a pair of Higgs bosons,  $pp \rightarrow X \rightarrow H H$, was simulated using \MGNLO[2.6.1]~\cite{Alwall:2014hca} at LO accuracy  with the  \NNPDF[2.3lo] PDF set. The event generator was interfaced with \HERWIG[7.1.3]~\cite{Bahr:2008pv,Bellm:2015jjp} to model the parton shower, hadronization and underlying event. No specific theoretical model for the phenomenology of the new particle is assumed in the process generation.
The mass of $X$  was varied between 251~\GeV{} and 1000~\GeV{} in the simulation, while its width was set to 10~\MeV{}. In total 25 $m_{X}$ mass hypotheses have been generated, corresponding to $m_{X}$ = 251, 260, 270, 280, 290, 300, 312.5, 325, 337.5, 350, 375, 400, 425, 450, 475, 500, 550, 600, 650, 700, 750, 800, 850, 900, 1000~\GeV{}.
The interference with nonresonant Higgs boson pair production was neglected.

Production of single Higgs bosons via ggF, VBF, $WH$, $ZH$ ($qq\to ZH$ and $gg\to ZH$), $t\bar{t}H$, $tH$ ($tHq$ and $tHW$), and $bbH$ was modeled using the same set of MC samples as described in Ref.~\cite{HIGG-2018-57}.
For single Higgs boson production, as well as both nonresonant and resonant di-Higgs production, a Higgs boson mass of 125.09~\GeV\ was assumed~\cite{HIGG-2014-14}. The analysis assumes a branching ratio of 0.227\% for the Higgs boson decay into two photons and a branching ratio of 58.2\% for the Higgs boson decay into two $b$-quarks~\cite{deFlorian:2016spz,DJOUADI199856}.
The inclusive cross sections of these processes are normalized to the most precise available theoretical values~\cite{deFlorian:2016spz}.

The $\gamma\gamma$+jets process was simulated with the \SHERPA[2.2.4]~\cite{Bothmann:2019yzt} generator. 
QCD NLO-accurate matrix elements for up to one parton, and LO-accurate matrix elements for up to three partons, were calculated with the Comix~\cite{Gleisberg:2008fv} and \OPENLOOPS~\cite{Buccioni:2019sur,Cascioli:2011va,Denner:2016kdg} libraries. These were calculated in the five-flavor scheme including $b$-quarks in the massless approximation and merged with the \SHERPA parton shower~\cite{Schumann:2007mg} using the \MEPSatNLO prescription~\cite{Hoeche:2011fd,Hoeche:2012yf} with a dynamic merging cut~\cite{Siegert:2016bre} of 10~\GeV. Within the parton shower, $b$-quarks were then treated as being massive.
Finally, events from $t\bar{t} \gamma \gamma$ processes were produced with \MGNLO\ in the four-flavor scheme~\cite{Alwall:2014hca}. The simulation samples used in the analysis are listed in Table~\ref{tab:single_higgs_samples_pyt}.
 
\begin{table}[ht!]
\centering
\caption{Summary of nominal Higgs boson pair signal samples and single-Higgs-boson background samples, split by production mode, and continuum background samples. The generator used in the simulation, the PDF set, and  set of tuned parameters (tune) are also provided.
\label{tab:single_higgs_samples_pyt}}
\resizebox{\columnwidth}{!}{
\begin{tabular}{ lllll }
\toprule
\midrule
Process & Generator    & PDF set    & Showering & Tune   \\
\midrule
\midrule
 
Nonresonant ggF $HH$  & \POWHEGBOX[v2]+FT~\cite{Nason:2009ai}~\cite{powheg-sm,powheg-kappa-lambda} & PDFLHC~\cite{Butterworth:2015oua}  &   \PYTHIA[8.2]~\cite{Sjostrand:2014zea} &  A14~\cite{ATL-PHYS-PUB-2014-021} \\
Nonresonant VBF $HH$  & \MGNLO~\cite{Alwall:2014hca} & \NNPDF[3.0nlo]~\cite{Ball:2014uwa} &  \PYTHIA[8.2] &  A14 \\
 
Resonant ggF $HH$  & \MGNLO& \NNPDF[2.3lo] &  \HERWIG[7.1.3]~\cite{Bahr:2008pv,Bellm:2015jjp} &   H7.1 - Default~\cite{Gieseke:2012ft}\\
 
\midrule
ggF $H$  & NNLOPS~\cite{Hamilton:2012np,Campbell:2012am,Hamilton:2012rf}~\cite{Bozzi:2005wk,deFlorian:2011xf}  & PDFLHC&   \PYTHIA[8.2]  &  AZNLO~\cite{Aad:2014xaa}      \\
VBF $H$ & \POWHEGBOX[v2]~\cite{Hamilton:2013fea,Nason:2004rx,Frixione:2007vw,Alioli:2010xd,Nason:2009ai,Mimasu:2015nqa,Campbell:2012am,Luisoni:2013kna,Hartanto:2015uka} & PDFLHC  &    \PYTHIA[8.2]     & AZNLO         \\
$WH$ &\POWHEGBOX[v2] & PDFLHC  &   \PYTHIA[8.2]      & AZNLO    \\
$qq\to ZH$ & \POWHEGBOX[v2] &  PDFLHC  &   \PYTHIA[8.2]      & AZNLO    \\
$gg\to ZH$ &  \POWHEGBOX[v2] & PDFLHC  &   \PYTHIA[8.2]      & AZNLO   \\
$t\bar{t}H$ & \POWHEGBOX[v2]~\cite{Frixione:2007nw,Nason:2004rx,Frixione:2007vw,Alioli:2010xd,Hartanto:2015uka} & \NNPDF[3.0nlo]  &   \PYTHIA[8.2]      &  A14  \\
$bbH$ &  \POWHEGBOX[v2] & \NNPDF[3.0nlo]  &  \PYTHIA[8.2]   & A14   \\
$tHq$ & \MGNLO &  \NNPDF[3.0nlo]  & \PYTHIA[8.2] &   A14    \\
$tHW$  & \MGNLO &  \NNPDF[3.0nlo]  & \PYTHIA[8.2] &  A14   \\
$\gamma\gamma+$jets &   \SHERPA[2.2.4]~\cite{Bothmann:2019yzt} & \NNPDF[3.0nnlo]  &  \SHERPA[2.2.4] & -- \\
$t\bar{t} \gamma \gamma$ & \MGNLO\  &  \NNPDF[2.3lo] & \PYTHIA[8.2] & -- \\
 
\bottomrule
\end{tabular}
}
\end{table}

Different pileup conditions from additional interactions the same and neighboring bunch crossings were simulated by overlaying the hard-scattering event with inelastic $pp$ events generated by \PYTHIA[8.186] using the \NNPDF[2.3lo] PDF set and the
A3 tune~\cite{ATL-PHYS-PUB-2016-017}. Differences between the simulated and observed distributions of the number of interactions per bunch crossing are corrected for by applying pileup scale factors to simulated events. A full simulation of the ATLAS detector~\cite{SOFT-2010-01} based on \GEANT~\cite{Agostinelli:2002hh} was used to reproduce the detector response to single-Higgs-boson processes. The continuum background and signal samples were processed by \atlfasttwo~\cite{ATLAS:1300517}, a fast simulation of the ATLAS detector response which was shown to be able to accurately simulate diphoton events.
% End of text imported from the .//sections/DataSimSamples.tex input file

% The next lines are included from the .//sections/object_event_selection.tex input file
\section{Object and event selections}

\subsection{Object selection}
\label{sec:object_sel}
Photons are reconstructed from topologically connected clusters~\cite{ATL-PHYS-PUB-2017-022} of energy deposits in the electromagnetic calorimeter in the region $|\eta| < 2.37$.
The transition region between the barrel and endcap electromagnetic calorimeters, $1.37 < |\eta| < 1.52$, is excluded.
Photon candidates matched to conversion vertices or tracks that are consistent with originating from photon conversions are classified as converted photons.
Those without a matched conversion vertex or track are classified as unconverted photons.
 
The calibration of the photon energy is based on a multivariate regression algorithm trained with MC samples, where the input variables are corrected with data-driven techniques. The calibrated energy is finally corrected by applying scale factors derived from $Z \rightarrow e^+ e^-$ events~\cite{EGAM-2018-01}.
The photon direction is reconstructed using the longitudinal (i.e.\ shower-depth) segmentation of the calorimeters and constrained to be compatible with the luminous region of the proton beam collisions.
Additionally, information about the position of the conversion vertex and about the tracks associated with the  conversion vertex is considered in the case of converted photons.
 
Events are required to have at least one reconstructed collision vertex, defined as a vertex associated with at least two tracks with transverse momentum ($\pt$) larger than $0.5$~\GeV.
The primary vertex is selected from the reconstructed collision vertices using a neural-network algorithm~\cite{HIGG-2013-08} based on the extrapolated photon trajectories and the tracks associated with each candidate vertex.
 
Photon identification is based on the lateral shower profile of the energy deposits in the first and second electromagnetic calorimeter layers and on the energy leakage fraction in the hadronic calorimeter.
It reduces the misidentification of hadronic jets containing large neutral components, primarily $\pi^0$ particles, which decay into a pair of highly collimated photons.
``Tight'' identification criteria, which are tuned for converted and unconverted photons separately, are applied~\cite{EGAM-2018-01}.
 
To further improve the rejection of misidentified photons, two isolation variables are defined to quantify the activity around a photon.
Calorimeter-based isolation $E^{\textrm{iso}}_{\textrm{T}}$ is defined as the sum of the transverse energy of topological clusters within a cone of size $\Delta R = 0.2$ around the photon, correcting for the energy of the photon candidate itself as well as for an average expected pileup contribution.
Track-based isolation $p_{\textrm{T}}^{\textrm{iso}}$ is defined as the scalar sum of the transverse momenta of all tracks with $\pT > 1$~\GeV{} that originate from the primary vertex 
and are within a cone of $\Delta R = 0.2$ around the photon. Isolated photons must have
$E^{\textrm{iso}}_{\textrm{T}} < 0.065\cdot E_\mathrm{T}$ and $p_{\textrm{T}}^{\textrm{iso}} < 0.05\cdot E_\mathrm{T}$ (the ``loose'' working point~\cite{EGAM-2018-01}), where $E_\mathrm{T}$ is the transverse energy of the photon.
For isolated photons with transverse momenta between $30~\GeV$ and $250~\GeV$, the identification efficiency for unconverted and converted photons ranges from 84\% to 98\%~\cite{EGAM-2018-01}.
 
Electrons are reconstructed from energy deposits measured in the electromagnetic calorimeter which are matched to ID tracks~\cite{EGAM-2018-01}.
They are required to satisfy $|\eta| < 2.47$, excluding the calorimeter transition region $1.37 < |\eta| < 1.52$, and have a transverse momentum $\pT > 10$~\GeV.
Electrons are required to satisfy a ``medium'' identification criterion based on the use of shower shape, track--cluster matching and TRT parameters in a likelihood-based algorithm~\cite{EGAM-2018-01}. 
Muons are reconstructed from high-quality tracks found in the MS~\cite{PERF-2015-10}.
A matching of these tracks to ID tracks is required in the region $|\eta| < 2.5$. Muons are required to have $|\eta| < 2.7$ and $\pT > 10$~\GeV, and to satisfy a ``medium'' identification criterion~\cite{MUON-2018-03}.
Both the electrons and muons are matched to the primary vertex via requirements on the tracks' longitudinal and transverse impact parameters, $|z_{0}|$ and $|d_{0}|$, respectively.
These requirements are $|z_{0}| \sin \theta < 0.5$~mm (where $\theta$ is the polar angle of the track) for electrons and muons and  $|d_{0}|/\sigma_{d_{0}} < 5\,(3)$ for electrons (muons).

Reconstructed jets are based on particle-flow objects built from noise-suppressed positive-energy topological clusters in the calorimeter and reconstructed tracks~\cite{PERF-2015-09}.
The anti-$\kt$ algorithm~\cite{Cacciari:2011ma,Cacciari:2008gp} with a radius parameter of $R=0.4$ is used.
They are required to have rapidity $|y|<4.4$ and $\pt>25~\gev$.
To further suppress jets produced in concurrent $pp$ interactions, each jet within the tracking acceptance of $|\eta|<2.4$, and with $\pT < 60~\GeV$, is required to satisfy the ``tight'' jet-vertex tagger~\cite{ATLAS-CONF-2014-018} criteria used to identify the jet as originating from the selected primary vertex of the event.
 
The flavor of jets is determined using a deep-learning neural network, DL1r.
The DL1r $b$-tagging is based on distinctive features of $b$-hadron decays in terms of the impact parameters
of the tracks and the displaced vertices reconstructed in the inner detector~\cite{FTAG-2018-01}.
The inputs of the DL1r network also include discriminating variables constructed by a recurrent neural network (RNNIP)~\cite{ATL-PHYS-PUB-2017-003}, which
exploits the spatial and kinematic correlations between tracks originating from the same $b$-hadron. For high-\pt jets, this
approach is found to give better performance~\cite{ATL-PHYS-PUB-2017-013} than a previously used multivariate technique~\cite{PERF-2012-04}. Operating points are
defined by a single selection value on the discriminant output distribution and are chosen to provide a specific
$b$-jet efficiency for an inclusive \ttbar{} MC sample.
The $b$-tagging requirements result in an efficiency of 77\% for jets containing $b$-hadrons ($b$-tagged jets), and the misidentification rate is 1/130 (1/4.9) for light-flavor (charm) jets~\cite{FTAG-2018-01}.
Scale factors are applied to the simulated events to correct for differences in $b$-tagging efficiency between data and simulation. These scale factors are measured as a function of jet \pt{} using a likelihood-based method in a sample highly enriched in \ttbar{} events~\cite{FTAG-2018-01}. Only jets with $|\eta|<2.5$ are considered for flavor-tagging.
 
The energy of $b$-tagged jets is corrected for the possible contribution of muons from semileptonic $b$-hadron decays. In addition, the undetected energy of the neutrinos and out-of-cone effects are corrected for with scale factors derived as a function of the $b$-jet \pt{} from a \ttbar{} MC sample. The two corrections together improve the resolution of the invariant mass of the two jets with the highest $b$-tagging score ($\mbb$) by about 20\%. The procedure closely follows the one used in Ref.~\cite{HIGG-2016-29}.
 
An overlap removal procedure is applied to avoid multiple usage of the same detector signals in the same event.
Photons are prioritized in this analysis, requiring the removal of electrons, muons and jets within $\Delta R = 0.4$ of a selected photon.
Next, jets within $\Delta R = 0.2$ of electrons are removed.
In the last step, electrons and muons within $\Delta R = 0.4$ of any jet are removed.
 
The missing transverse momentum \etmiss{} in an event is calculated as the magnitude of the negative vectorial sum of the transverse momenta of all selected and calibrated physics objects that can be matched to the primary vertex. A component called the ``soft term'' is calculated from the residual tracks that originate from the primary vertex but are not associated with any other object and is added to the \etmiss calculation~\cite{PERF-2016-07}.

\subsection{Event selection}
\label{sec:event_selection}
Both the nonresonant and resonant $HH$ searches employ multivariate analysis techniques to select events.
Events are selected if they satisfy a common set of preselection requirements; they are then required to fulfill different requirements for the nonresonant search and the resonant search. Both analyses employ a fit of the diphoton invariant mass distribution to extract the $HH$ signal contribution.
 
\subsubsection{Common preselection}
\label{subsec:preselection}
For both the nonresonant and resonant analyses, events are selected using diphoton triggers requiring two reconstructed photon candidates with minimum transverse energies of 35~\GeV\ for the leading photon and 25~\GeV\ for the subleading photon, where leading (subleading) refers to the photon candidate with the highest (second-highest) transverse energy~\cite{Aad:2019wsl}.
The triggers used in 2015 and 2016 required both photons to satisfy a Loose photon identification criterion, while the Medium criterion~\cite{EGAM-2018-01} was adopted in 2017--2018 to cope with the increased $pp$ interaction rate.
Once the full diphoton event selection described in this Section is applied, the average trigger efficiency for $H\to\gamma\gamma$ events is found to be greater than 99\% for the 2015--2016 data-taking period, and greater than 98\% for the 2017--2018 data-taking period.
 
On top of the trigger requirements, events are selected if:
\begin{itemize}
\item At least two photons satisfy the object selection criteria detailed in Section~\ref{sec:object_sel}.
\item The diphoton invariant mass, built with the two leading photons, satisfies $105 < m_{\gamma\gamma} < 160$~\GeV.
\item The leading (subleading) photon \pT is larger than 35\% (25\%) of the mass of the diphoton system.
\item Exactly two $b$-tagged jets are present. In order to remain statistically independent of the ATLAS search for $HH \rightarrow b\bar{b}b\bar{b}$~\cite{EXOT-2016-31}, any event with more than two $b$-jets passing the 77\% efficient working point is rejected.
\item No electrons or muons are present.
\item Fewer than six central ($|\eta|$ < 2.5) jets are present. This helps to reject $t\bar{t}H$ events where the top quarks decay hadronically.
\end{itemize}
 
The acceptance times efficiency of the common preselection for the SM ggF $HH$ simulation sample is 14\% and it is 8.5\% (14\%) for a resonant scalar particle with $m_{X}=300~\GeV$ ($m_{X}=500~\GeV$).
Multivariate techniques are then used to target the nonresonant \ggF\ production mode or the resonant production mode of Higgs boson pairs.
For both the nonresonant and resonant analyses, the resulting categories, each based on a boosted decision tree (BDT), must have at least nine continuum background events in the data sideband region.
This guarantees that the final selection on the data sideband retains enough events to perform a fit to the distribution of the diphoton invariant mass, $\myy$.
The data sideband region is defined as the range $105 < m_{\gamma\gamma} < 160$~\GeV, excluding $120 < m_{\gamma\gamma} < 130$~\GeV.
 
The invariant mass of the diphoton plus b-tagged jets system, $\myybb^{*}$, is defined as $\myybb^{*} = m_{\yybb} - m_{\bb} - m_{\gamma\gamma} + 250~\GeV$ (where 250~\GeV\ is about twice the Higgs boson mass value and $m_{\bb}$ is the invariant mass of the two jets with the highest $b$-tagging score). It is used to implement selection criteria for both the nonresonant and resonant analyses.
Figure~\ref{fig:mc-myybb} shows that, compared with $\myybb$, the $\myybb^{*}$ variable improves the four-object mass resolution, particularly for resonant signal particles decaying into a pair of Higgs bosons, due to detector resolution effects canceling out.
 
\begin{figure}[!htb]
\centering
\includegraphics[width=0.55\textwidth]{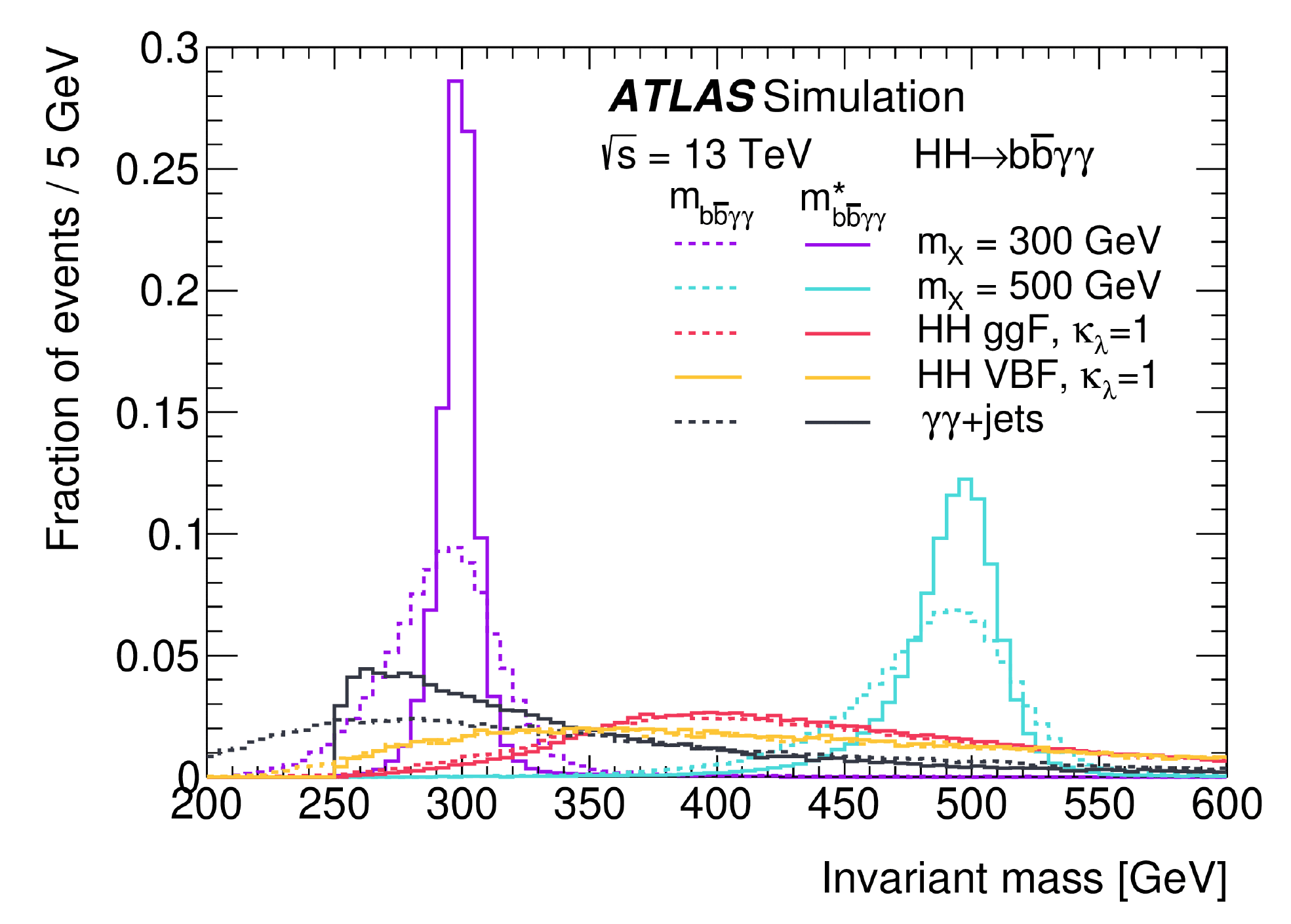}
\caption{Reconstructed four-body mass for $m_X = 300~\GeV$ and $m_X = 500~\GeV$ resonant signal benchmarks, for the SM $HH$ production processes and for the $\gamma\gamma$+jets background. Dashed lines represent the distribution of $\myybb$ while solid lines represent the distribution of $\myybb^{*}$, defined in Section~\ref{subsec:preselection}. Distributions are normalized to unit area.}
\label{fig:mc-myybb}
\end{figure}

\subsubsection{Nonresonant selection}
\label{subsec:BDTnon_resonant_selection}
Following the preselection, events are divided into two regions using the value of the $\myybb^{*}$ variable.
A high mass region, with $\myybb^{*} > 350~\GeV$, targets the SM signal ($\klambda=1$), while a low mass region, with $\myybb^{*} < 350~\GeV$, is used to retain sensitivity for BSM signals ($\klambda=10$). The dependence of $\myybb^{*}$ on $\klambda$ can be seen in Figure~\ref{fig:mx-vs-kl}.

\begin{figure}[!htb]
\centering
\subfloat[ggF $HH$ production mode]{
\includegraphics[width=0.45\textwidth]{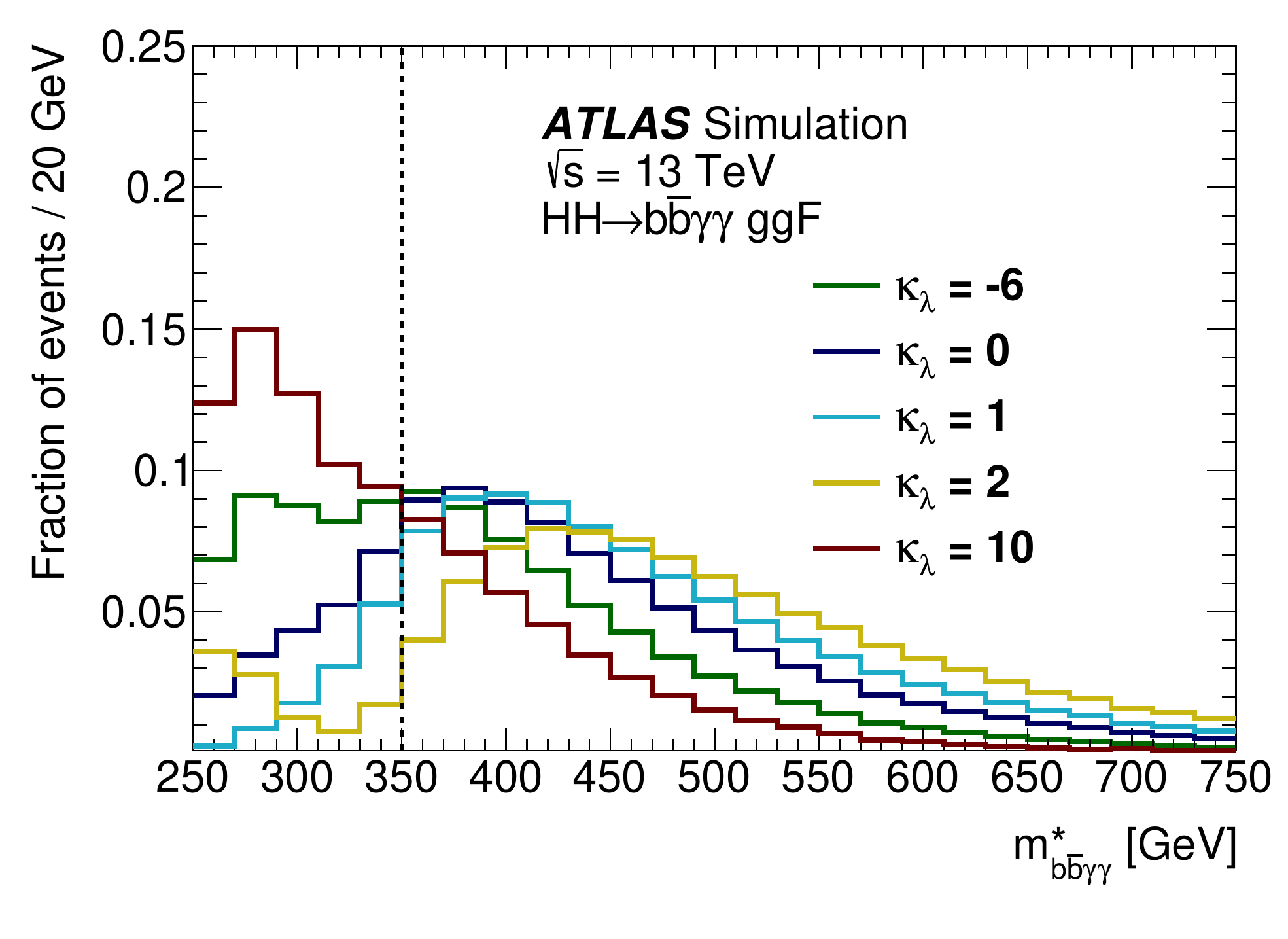}}
\quad
\subfloat[VBF $HH$ production mode]{
\includegraphics[width=0.45\textwidth]{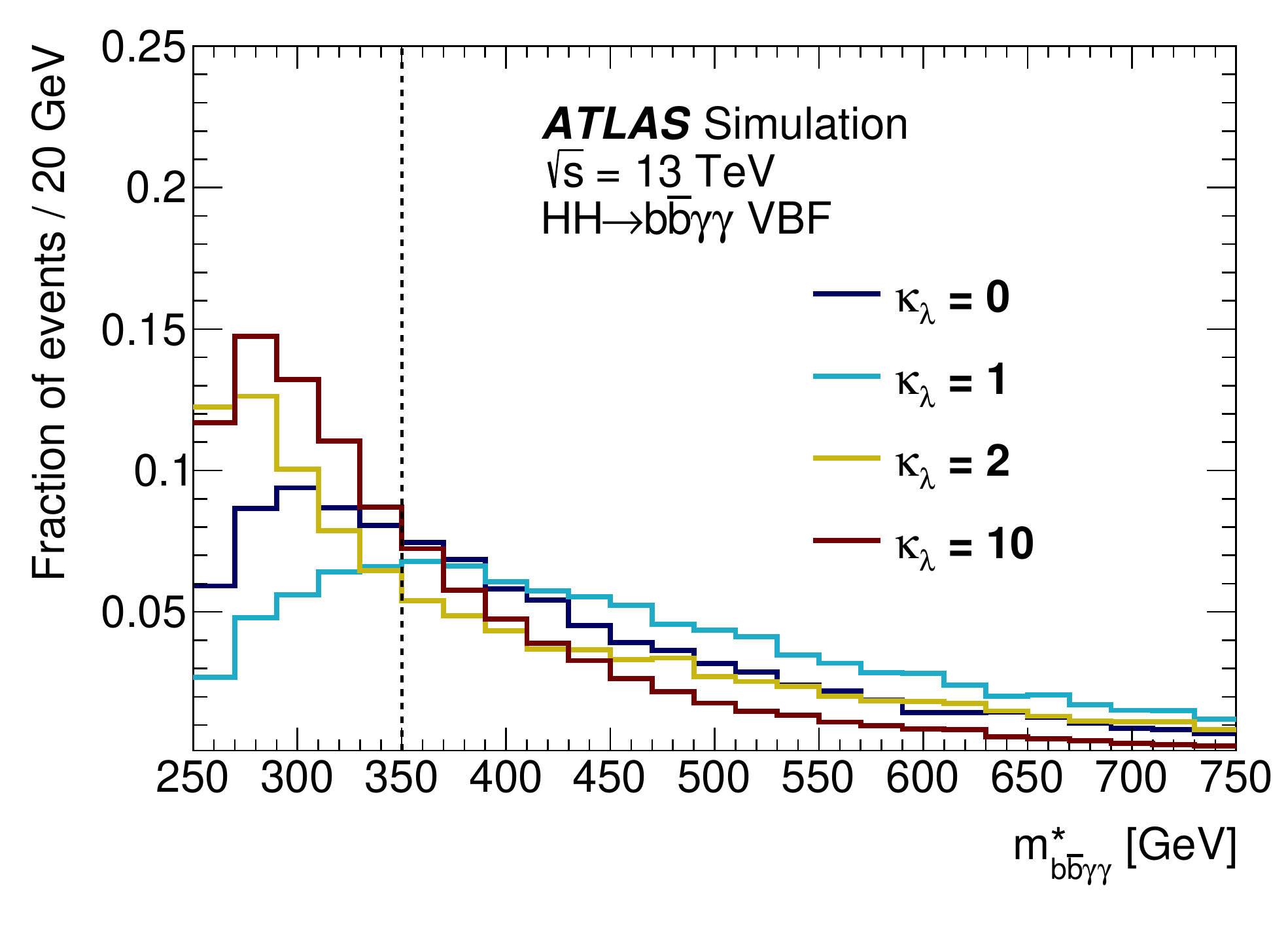}}
\caption{The $\myybb^{*}$ distributions after the common preselection for (a) nonresonant ggF $HH$ and (b) VBF $HH$ signals with several $\klambda$ values. The value of $\myybb^{*} = 350~\GeV$ is chosen as the boundary between categories targeting the SM and BSM $\klambda$ signals.}
\label{fig:mx-vs-kl}
\end{figure}

In each mass region, a dedicated BDT is trained using XGBoost~\cite{Chen:2016:XST:2939672.2939785} to discriminate between a benchmark $HH$ signal and a combination of $\yy$, $t\bar{t}H$, $ggH$, and $ZH$ simulated backgrounds. In the high mass region, the SM $HH$ sample is used as signal, while in the low mass region, the $\klambda = 10$ sample is used as signal.

The BDT input variables are summarized in Table~\ref{tab:BDT_nonRes_input}. Identical variable sets are used for high mass and low mass categories.
The BDT combines several input variables that exploit the different kinematic properties of signal and background events, as well as the $b$-tagging information.
Observables based on the kinematic properties of the reconstructed photons, such as the leading and subleading photon's angular information, and the transverse momentum of the diphoton system divided by its invariant mass, are combined with jet-based information.
The ``single topness'' variable ($\chi_{Wt}$) is also used. It is defined as
 
\begin{equation}
\label{eq:topness}
\chi_{Wt} = \text{min}\sqrt{{ \left(\frac{m_{j_1j_2} - m_W}{m_W} \right)}^2 + { \left(\frac{m_{j_1j_2j_3} - m_t}{m_t} \right)}^2},
\end{equation}
where the minimum is taken over all combinations of three jets in the event (with no requirements on $b$-tagging status), $m_W = 80~\GeV$, and $m_t = 173~\GeV$.
Among the input variables in Table~\ref{tab:BDT_nonRes_input}, $\mbb$ and $H_{\text{T}}$ show the highest discriminating power against the $\gamma \gamma$+jets continuum background.
Particular care was taken to ensure that the BDT event selection does not lead to biases in the \myy{} background distribution.
Variables which have a strong correlation with the diphoton invariant mass are avoided in the training in order to prevent the BDT event selection from biasing the \myy{} background distribution.
To this end, the transverse momentum values of the photons are divided by \myy{} before being used as BDT input variables. A check for potential biases in the \myy{} background distribution is described in Section~\ref{sec:backpar}.
The BDT score distributions in the low mass and high mass regions are shown in Figure~\ref{fig:nonres-ggf-bdt} for events passing the common preselection. In each mass region, two categories based on the BDT score are defined. The boundaries of the categories are chosen by maximizing the combined number-counting significance~\cite{Cowan:2010js} using signal and background yields in the mass window $120 < m_{\gamma\gamma} < 130$~\GeV\ in the chosen categories.
The four resulting BDT categories are defined in Table~\ref{tab:nonres-cat}.
 
\begin{table}[!htb]
\centering
\renewcommand{\arraystretch}{.8}
\caption{Variables used in the BDT for the nonresonant analysis.
All vectors in the event are rotated so that the leading photon $\phi$ is equal to zero, while their relative  azimuthal angular differences are kept unchanged.
\label{tab:BDT_nonRes_input}}.
\begin{tabular}{ll}
\toprule
\midrule
Variable & Definition  \\
\midrule
\midrule
\multicolumn{2}{l}{Photon-related kinematic variables} \\
\midrule
$\pt$/$m_{\yy}$    & \begin{minipage}[l]{0.55\textwidth} Transverse momentum of each of the two photons divided by the diphoton invariant mass $m_{\yy}$  \end{minipage} \\[2ex]
$\eta$ and $\phi$    & \begin{minipage}[l]{0.55\textwidth} Pseudorapidity and azimuthal angle of the leading and subleading photon  \end{minipage} \\[2ex]
\midrule
\midrule
\multicolumn{2}{l}{Jet-related kinematic variables} \\
\midrule
$b$-tag status & \begin{minipage}[l]{0.55\textwidth} Tightest fixed $b$-tag working point (60\%, 70\%, or 77\%) that the jet passes \end{minipage} \\[2ex]
$\pt$, $\eta$ and $\phi$   & \begin{minipage}[l]{0.55\textwidth} Transverse momentum, pseudorapidity and azimuthal \\ angle of the two jets with the highest $b$-tagging score \end{minipage}\\[2ex]
$\pt^{b\bar{b}}$, $\eta_{b\bar{b}}$ and $\phi_{b\bar{b}}$   & \begin{minipage}[l]{0.55\textwidth} Transverse momentum, pseudorapidity and azimuthal \\ angle of the $b$-tagged jets system \end{minipage}\\[2ex]
$m_{b\bar{b}}$   & \begin{minipage}[l]{0.55\textwidth} Invariant mass of the two jets with the highest $b$-tagging score \end{minipage} \\[2ex]
$H_{\text{T}}$ & \begin{minipage}[l]{0.55\textwidth} Scalar sum of the $\pt$ of the jets in the event \end{minipage}  \\[2ex]
Single topness & \begin{minipage}[l]{0.55\textwidth} For the definition, see Eq.~(\ref{eq:topness}) \end{minipage} \\[2ex]
\midrule
\midrule
\multicolumn{2}{l}{Missing transverse momentum variables} \\
\midrule
\etmiss{} and $\phi^{\text{miss}}$ & \begin{minipage}[l]{0.55\textwidth} Missing transverse momentum and its azimuthal angle \end{minipage}  \\[2ex]
\bottomrule
\end{tabular}
\end{table}
 
\begin{figure}[!htb]
\centering
\subfloat[Low mass region]{\includegraphics[width=0.45\textwidth]{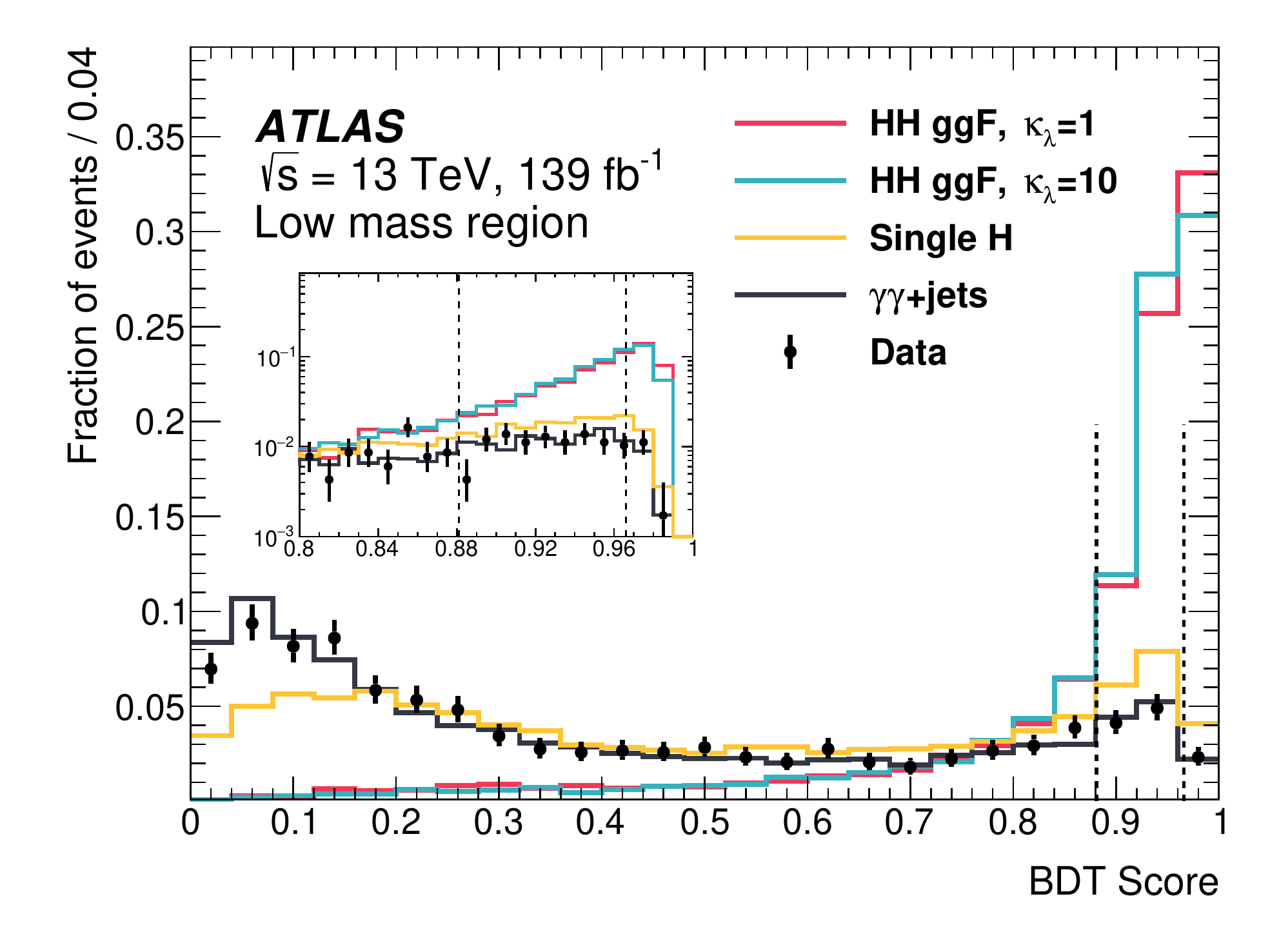}}
\subfloat[High mass region]{\includegraphics[width=0.45\textwidth]{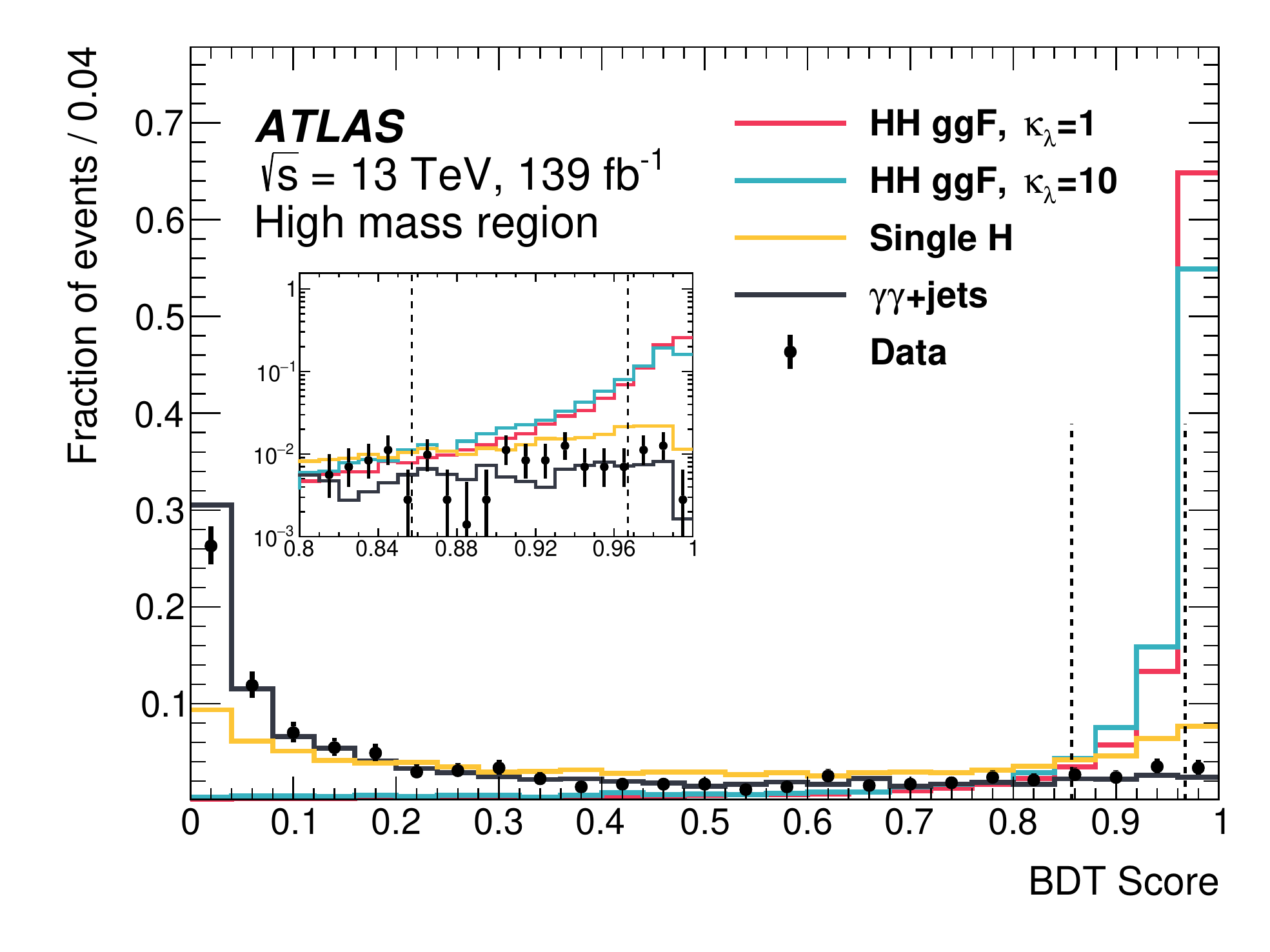}}
\caption{The BDT distribution of the di-Higgs ggF signal for two different values of $k_{\lambda}$ and the main backgrounds in the (a) low mass region and (b) high mass region. Distributions are normalized to unit area. The dashed lines denote the category boundaries. Events with a BDT score below $0.881$ in the low mass region or below $0.857$ in the high mass region are discarded.}
\label{fig:nonres-ggf-bdt}
\end{figure}

\begin{table}[!htb]
\renewcommand{\arraystretch}{1.}
\begin{center}
\caption{Definition of the categories used in the $HH$ nonresonant search. Before entering the BDT-based categories, events are required to satisfy the common preselection.\label{tab:nonres-cat}}
\begin{tabular}{ll}
\toprule
\midrule
Category & Selection criteria \\
\midrule
\midrule
High mass BDT tight & $\myybb^{*} \geq 350~\GeV$, $\text{BDT\,score} \in [0.967, 1] $ \\
High mass BDT loose & $\myybb^{*} \geq 350~\GeV$, $\text{BDT\,score} \in [0.857, 0.967] $ \\
Low mass BDT tight & $\myybb^{*} < 350~\GeV$, $\text{BDT\,score} \in [0.966, 1] $ \\
Low mass BDT loose & $\myybb^{*} < 350~\GeV$, $\text{BDT\,score} \in [0.881, 0.966] $ \\
\bottomrule
\end{tabular}
\end{center}
\end{table}
 
\subsubsection{Resonant selection}
The resonant search uses a multivariate analysis based on a BDT technique.
A potential limitation of a BDT-based selection is the low number of background events for higher resonance masses. To overcome this limitation, BDTs are trained jointly for all resonance masses hypotheses and background. The signal events corresponding to the different mass hypotheses are combined.
The signal events are reweighted event-by-event to match the $\myybb^{*}$ distribution of the background events, such that the training is independent of the resonant signal mass hypothesis, but it still retains information of the correlation with the rest of the event variables. The procedure allows to reduce the fluctuations of the BDT performance between nearby mass points. It was checked that the procedure provides similar or better performance than employing separated BDT trainings for each individual signal mass hypothesis.
 
Using the TMVA toolkit~\cite{TMVA}, two BDTs are trained to better separate the signal from backgrounds of different nature: the $\gamma\gamma$ and $t\bar{t}\gamma\gamma$ backgrounds ($\textrm{BDT}_{\gamma\gamma}$) and the single-Higgs-boson background ($\textrm{BDT}_{\mathrm{Single} H}$), where the $ZH$ and $t\bar{t}H$ processes produce the dominant resonant backgrounds. The nonresonant $HH$ process is not included in the training of the BDTs.
A complete list of the variables used for the BDT training is given in Table~\ref{tab:BDT_Res_input}.
The \etmiss{} information is used in the training because it is useful in rejecting the single-Higgs-boson background ($t\bar{t}H$ in particular) and the $t\bar{t} \gamma \gamma$ background.
 
\begin{table}[!htb]
\centering
\renewcommand{\arraystretch}{.8}
\caption{Variables used in the BDT for the resonant analysis.
For variables depending on $b$-tagged jets, only jets $b$-tagged using the 77\% working point are considered as described in Section~\ref{sec:object_sel}.   \label{tab:BDT_Res_input}}
\begin{tabular}{ll}
\toprule
\midrule
Variable & Definition  \\
\midrule
\midrule
\multicolumn{2}{l}{Photon-related kinematic variables} \\
\midrule
$p_{\mathrm{T}}^{\gamma \gamma}$, $y^{\gamma \gamma}$  & \begin{minipage}[l]{0.55\textwidth} Transverse momentum and rapidity of the diphoton system  \end{minipage} \\[2ex]
$\Delta \phi_{\gamma \gamma}$ and $\Delta R_{\gamma \gamma}$  & \begin{minipage}[l]{0.55\textwidth} Azimuthal angle and $\Delta R$ between the two photons  \end{minipage} \\[2ex]
\midrule
\midrule
\multicolumn{2}{l}{Jet-related kinematic variables} \\
\midrule
$m_{b \bar{b}}$, $p_{\mathrm{T}}^{b \bar{b}}$ and $y_{b \bar{b}}$  & \begin{minipage}[l]{0.55\textwidth} Invariant mass, transverse momentum and rapidity of the $b$-tagged jets system  \end{minipage} \\[2.5ex]
$\Delta \phi_{b \bar{b}}$  and $\Delta R_{b \bar{b}}$  & \begin{minipage}[l]{0.55\textwidth} Azimuthal angle and $\Delta R$ between the two $b$-tagged jets  \end{minipage} \\[2.5ex]
$N_{\text{jets}}$ and $N_{b\text{-jets}}$   & \begin{minipage}[l]{0.55\textwidth} Number of jets  and number of $b$-tagged jets\end{minipage} \\[2ex]
$H_{\text{T}}$ & \begin{minipage}[l]{0.55\textwidth} Scalar sum of the $\pt$ of the jets in the event \end{minipage}  \\[2ex]
\midrule
\midrule
\multicolumn{2}{l}{Diphoton+dijet-related kinematic variables} \\
\midrule
$\myybb^{*}$    & \begin{minipage}[l]{0.55\textwidth} Invariant mass of the diphoton plus $b$-tagged jets system  \end{minipage} \\[2.5ex]
$\Delta y_{\gamma \gamma, b \bar{b}}$, $\Delta \phi_{\gamma \gamma, b  \bar{b}}$ and $\Delta R_{\gamma \gamma, b \bar{b}}$  & \begin{minipage}[l]{0.55\textwidth} Distance in rapidity, azimuthal angle and  $\Delta R$ between the diphoton and the $b$-tagged jets system  \end{minipage} \\[2ex]
\midrule
\midrule
\multicolumn{2}{l}{Missing transverse momentum  variables} \\
\midrule
\etmiss{} & \begin{minipage}[l]{0.55\textwidth} Missing transverse momentum \end{minipage}  \\[2ex]
\bottomrule
\end{tabular}
\end{table}
 
The combined BDT score of an event is obtained by combining the two BDT scores in quadrature, as shown in Eq.~(\ref{eqn:qudrature-sum-scores}):
 
\begin{equation}
\label{eqn:qudrature-sum-scores}
\textrm{BDT}_{\textrm{tot}} = \frac{1}{\sqrt{C_1^2+C_2^2}}\sqrt{C_1^2\left(\frac{\textrm{BDT}_{\gamma\gamma}+1}{2}\right)^2+C_2^2\left(\frac{\textrm{BDT}_{\mathrm{Single} H}+1}{2}\right)^2}.
\end{equation}
 
The two coefficients $C_1$ and $C_2$ ($C_2 = 1-C_1$) and $\textrm{BDT}_{\textrm{tot}}$ take values in the range [0,1]. Only events passing a minimum requirement on the value of $\textrm{BDT}_{\textrm{tot}}$ are considered in the analysis. To reduce the effect of limited statistics of the Monte Carlo samples in the determination the minimum value required for $\textrm{BDT}_{\textrm{tot}}$, and of the  $C_1$ and  $C_2$ coefficients, a two-stage procedure is used. The values of the $C_1$ and $C_2$ as well as the BDT cut value are scanned twice in order to maximize the significance. This two-stage optimization procedure first finds the maximum significance that can be achieved for each resonance mass point independently, leading to different coefficient values and $\textrm{BDT}_{\textrm{tot}}$ requirement values for the various mass points. A second scan is done to select all coefficients providing a significance within 5\% of the maximum value, for each of the resonance mass values.
From those possible combinations a common $C_1$ coefficient is sought (resulting in $C_1 = 0.65$) across all the resonances so that the selection will have common coefficients for all resonance mass points, but different  $\textrm{BDT}_{\textrm{tot}}$ values. 
For each of the resonance mass hypotheses, a requirement is set on the $\myybb^{*}$ value to select events within $\pm2\sigma$ of the expected mean value for signal events (where $\sigma$ is the standard deviation parameter of the Crystal Ball function that best fits the $\myybb^{*}$ distribution). In the case of the 900~\GeV{} and 1000~\GeV{} mass hypotheses, the requirement is relaxed to  $\pm4\sigma$ to increase the number of events used for the signal extraction.
The $\textrm{BDT}_{\textrm{tot}}$ distributions are shown in Figure~\ref{fig:res-bdt} for two signal hypotheses ($m_X = 300~\GeV$ and $500~\GeV$).
 
\begin{figure}[htbp]
\centering
\subfloat[]{\includegraphics[width=0.45\textwidth]{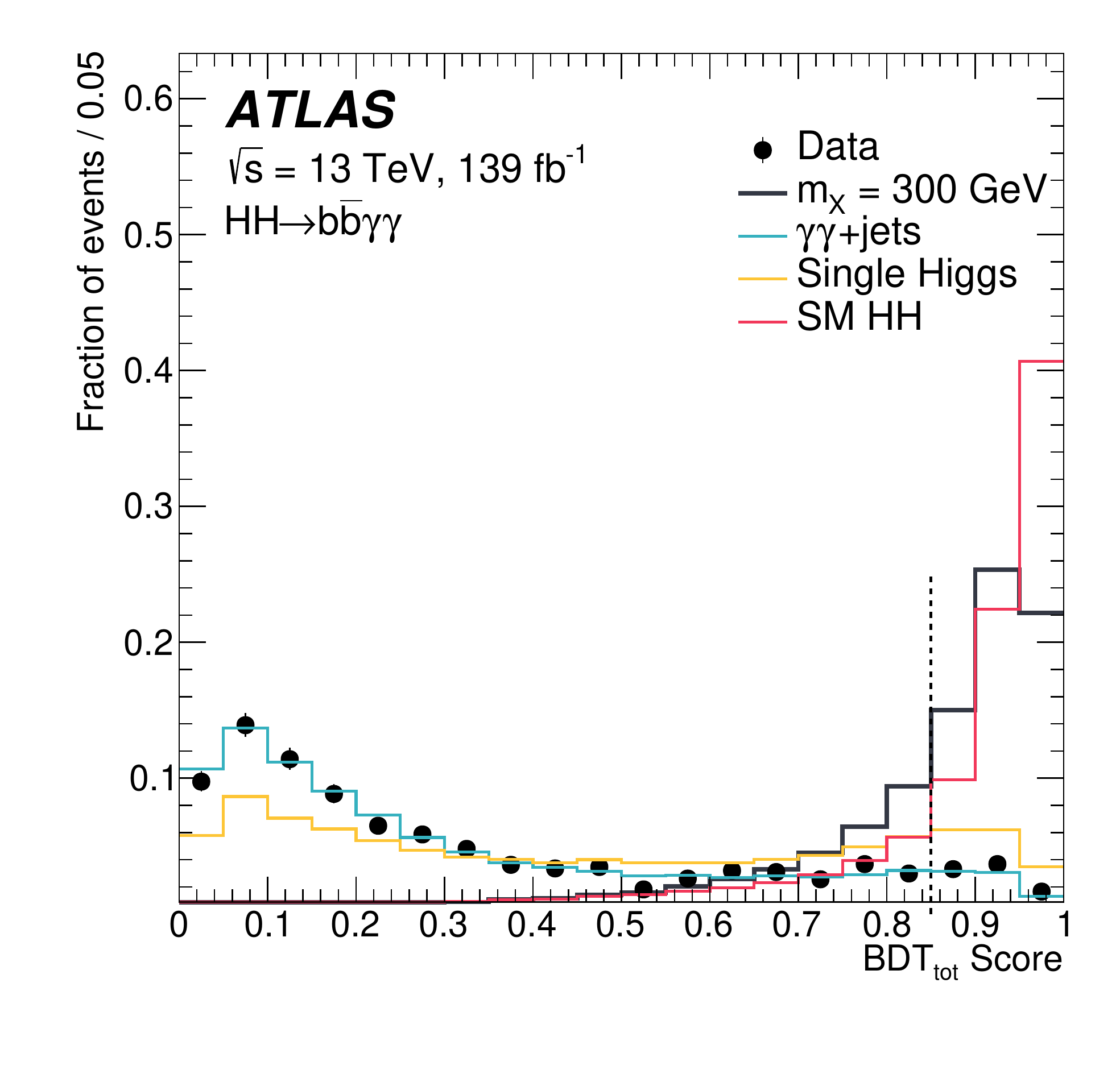}}
\subfloat[]{\includegraphics[width=0.45\textwidth]{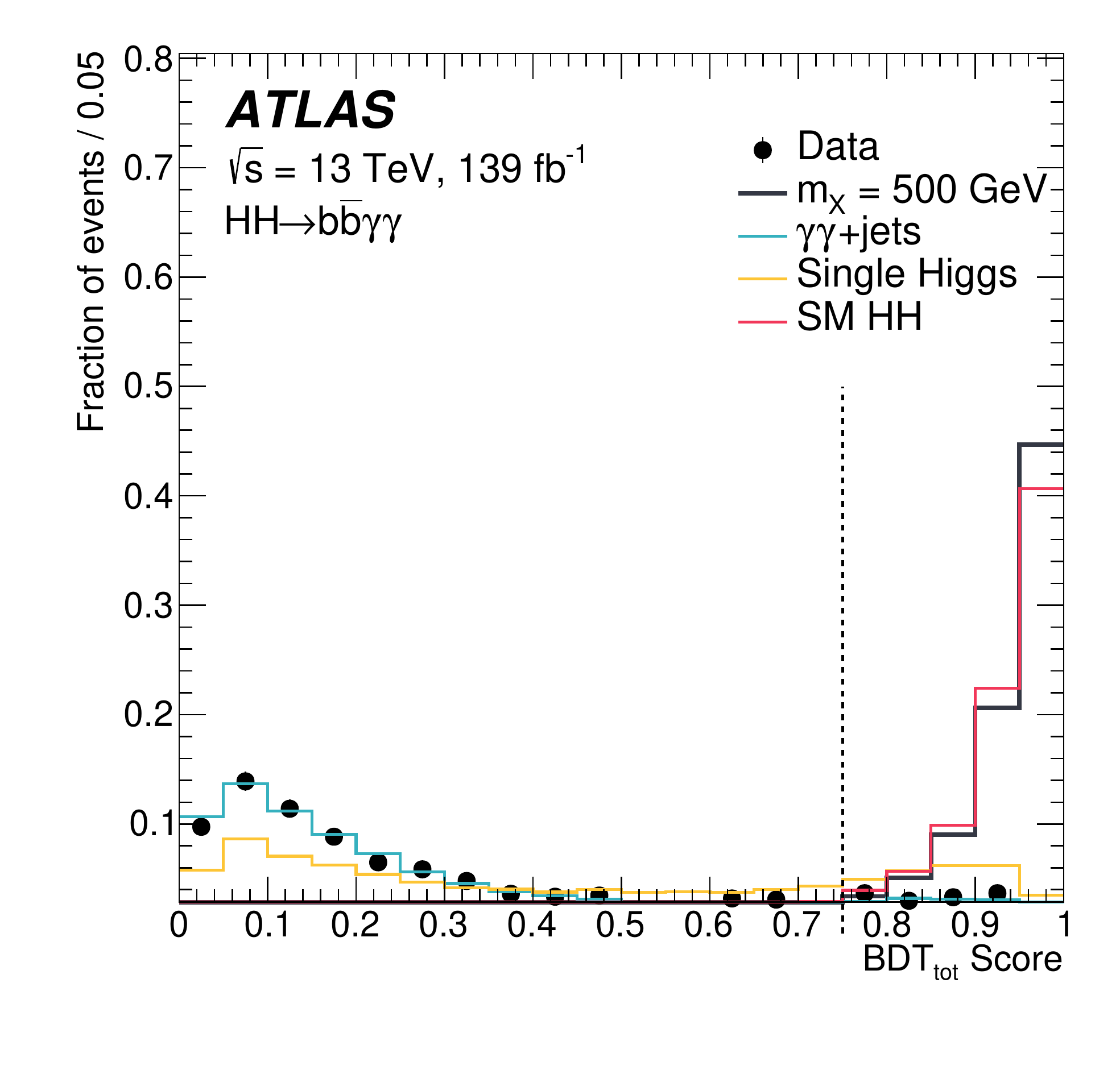}}
\caption{The $\textrm{BDT}_{\textrm{tot}}$ score for the benchmark signals ((a) $m_X = 300$~\GeV\ and (b) $m_X = 500$~\GeV) and for the main backgrounds. Distributions are normalized to unit area. The dashed lines denote the event selection thresholds. Events with a $\textrm{BDT}_{\textrm{tot}}$ score below $0.85$ for $m_X = 300$~\GeV\ or below $0.75$ for  $m_X = 500$~\GeV\ are discarded.}
\label{fig:res-bdt}
\end{figure}
 
\subsection{Comparison of data and predictions}
\label{sec:background_diphoton}
 
The analysis selection described in Section~\ref{sec:event_selection} requires two ``tight'' photons and this region is mainly composed of $\gamma \gamma$, $\gamma$--jet, and dijet events, where either one or two jets are misidentified as photons.
The fractional contribution of each component can be estimated using a data-driven approach~\cite{STDM-2012-05} based on the photon identification and isolation distributions from genuine and misidentified photons.
After the common preselection, $(85 \pm 3)$\% of sideband events are genuine diphoton events, with the remaining $(15 \pm 4)$\% consisting of $\gamma$--jet events and a negligible number of dijet events. The uncertainties in the above fractions include both the statistical and systematic uncertainties, where the systematic uncertainty is estimated by using different photon identification criteria.
In the BDT-based categories, the proportion of genuine diphoton events increases but the method suffers from a low event count for both the nonresonant and resonant cases. 
 
Figure~\ref{fig:dataMC_preselection} shows the level of agreement between data and the background prediction for the \myy and $\myybb^{*}$ distributions, after the common preselection.
The continuum background is scaled by the $\gamma \gamma$, $\gamma$--jet, and dijet fractions and normalized to the data sideband.
The $\gamma \gamma$+jets continuum background is further divided according to the flavors of the two jets (for example $b\bar{b}$ or other jets). This decomposition is taken directly from the proportions predicted by the \SHERPA event generator, as described in Section~\ref{sec:simulation}, and it is shown for illustration.
Figures~\ref{fig:dataMC_myy_nonRes} and~\ref{fig:dataMC_myy_Res} show the \myy distribution after the nonresonant and resonant BDT categorization and for two benchmark mass points $m_X = 300$~\GeV{} and $m_X = 500$~\GeV.
The figures illustrate the signal and background composition. The background distributions shown in these figures are not directly used to model the background processes in the analysis workflow  explained in Section~\ref{sec:sigbackpar}.
 
\begin{figure}
\centering
\subfloat[]{
\includegraphics[width=0.4\textwidth]{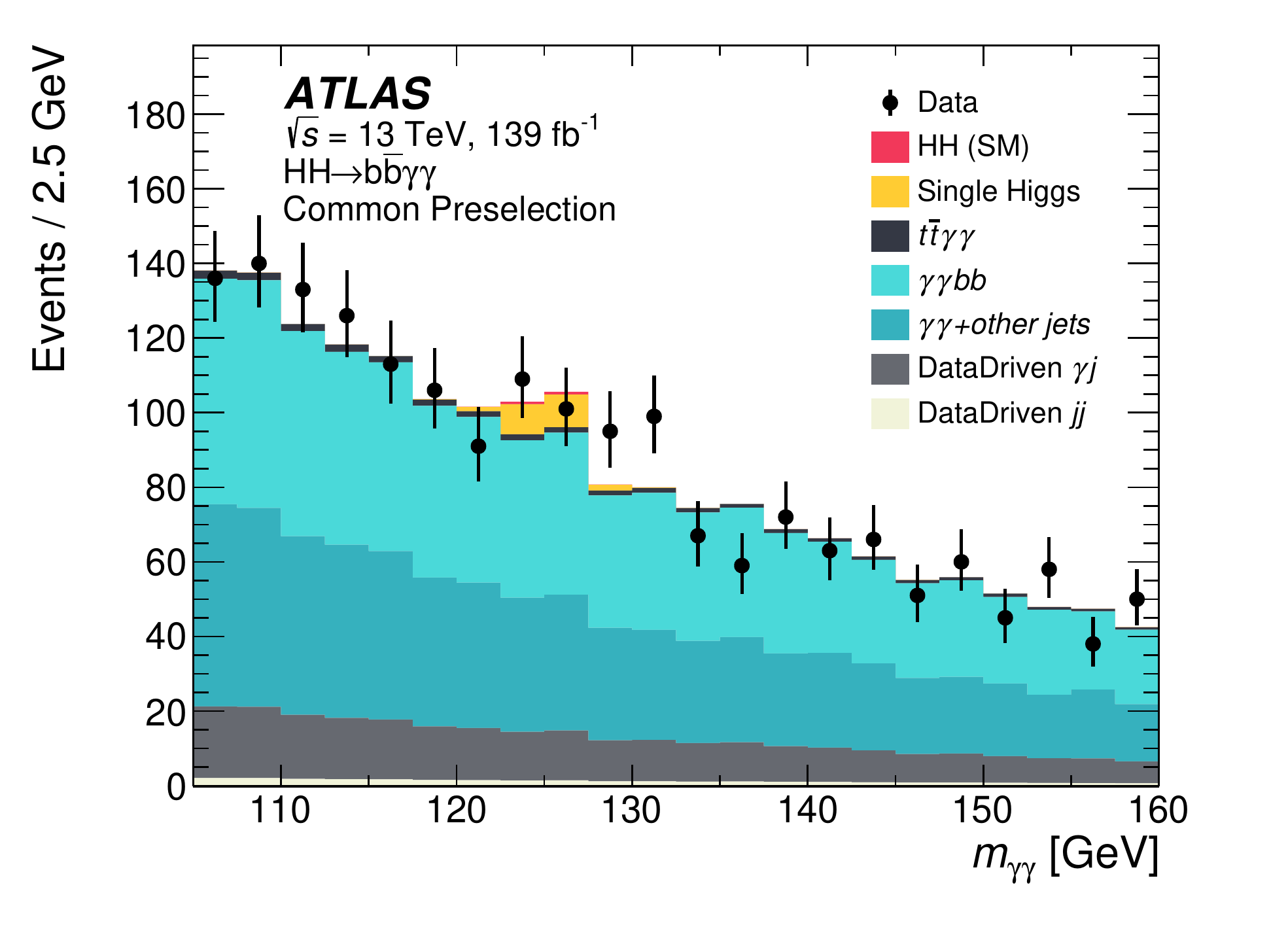}}
\subfloat[]{
\includegraphics[width=0.4\textwidth]{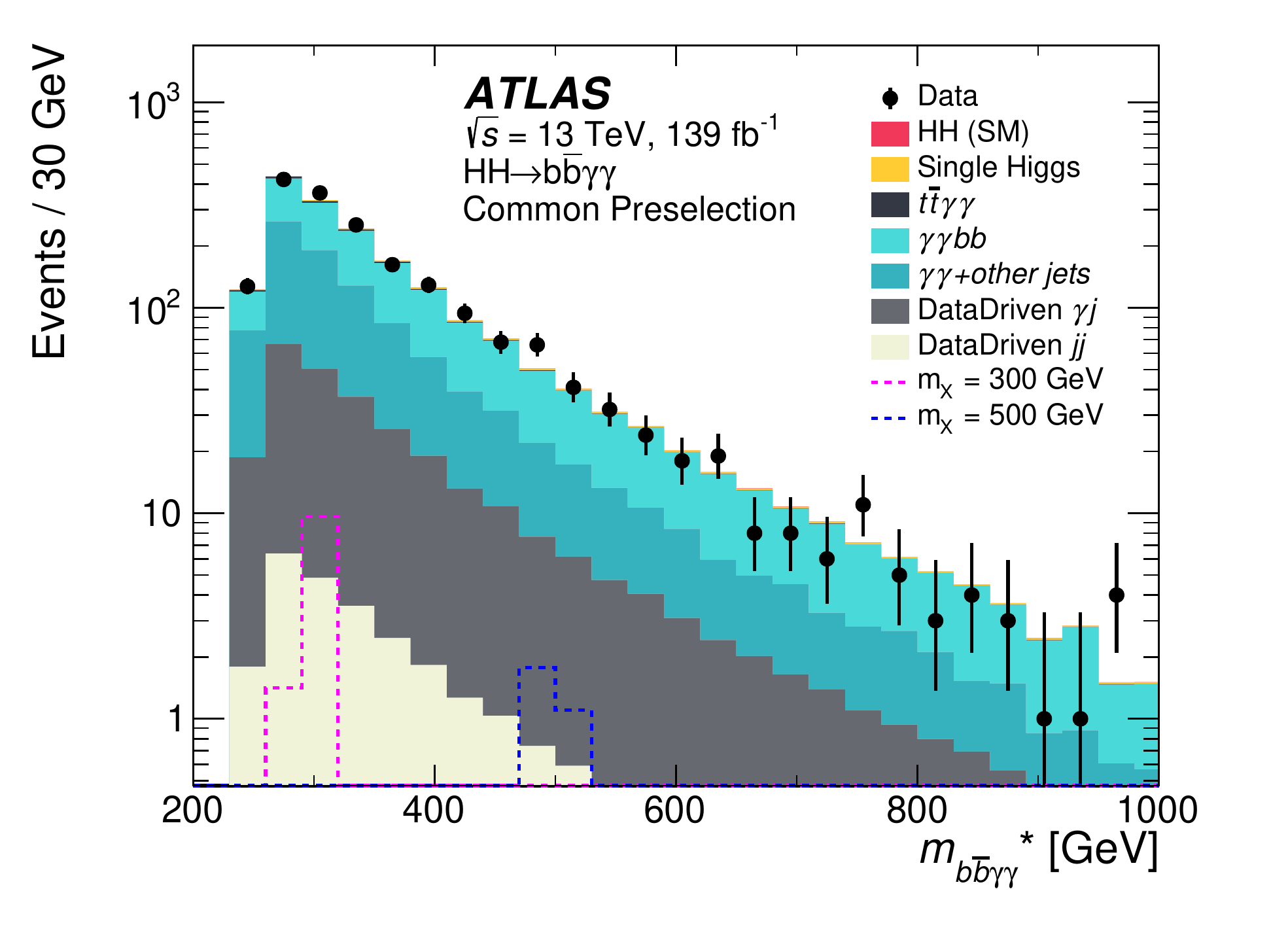}}
\caption{Distributions of (a) \myy{} and (b) $\myybb^{*}$ for events satisfying the common preselection criteria.
The data-derived fractions of nonresonant $\gamma \gamma$, $\gamma$--jet or jet--$\gamma$, and dijet background are applied and the total background is normalized to the data sideband.
The scalar resonance signal is scaled to a total production cross section $\sigma(pp\to X \to HH)=370$~fb  for $m_{X}=300~\GeV$ or $\sigma(pp\to X \to HH)=67$~fb for $m_{X}=500~\GeV$, corresponding to the expected exclusion limits presented in Section~\ref{sec:results_resonant}.}
\label{fig:dataMC_preselection}
\end{figure}
 
\begin{figure}
\centering
\subfloat[High mass BDT tight selection]{\includegraphics[width=0.4\textwidth]{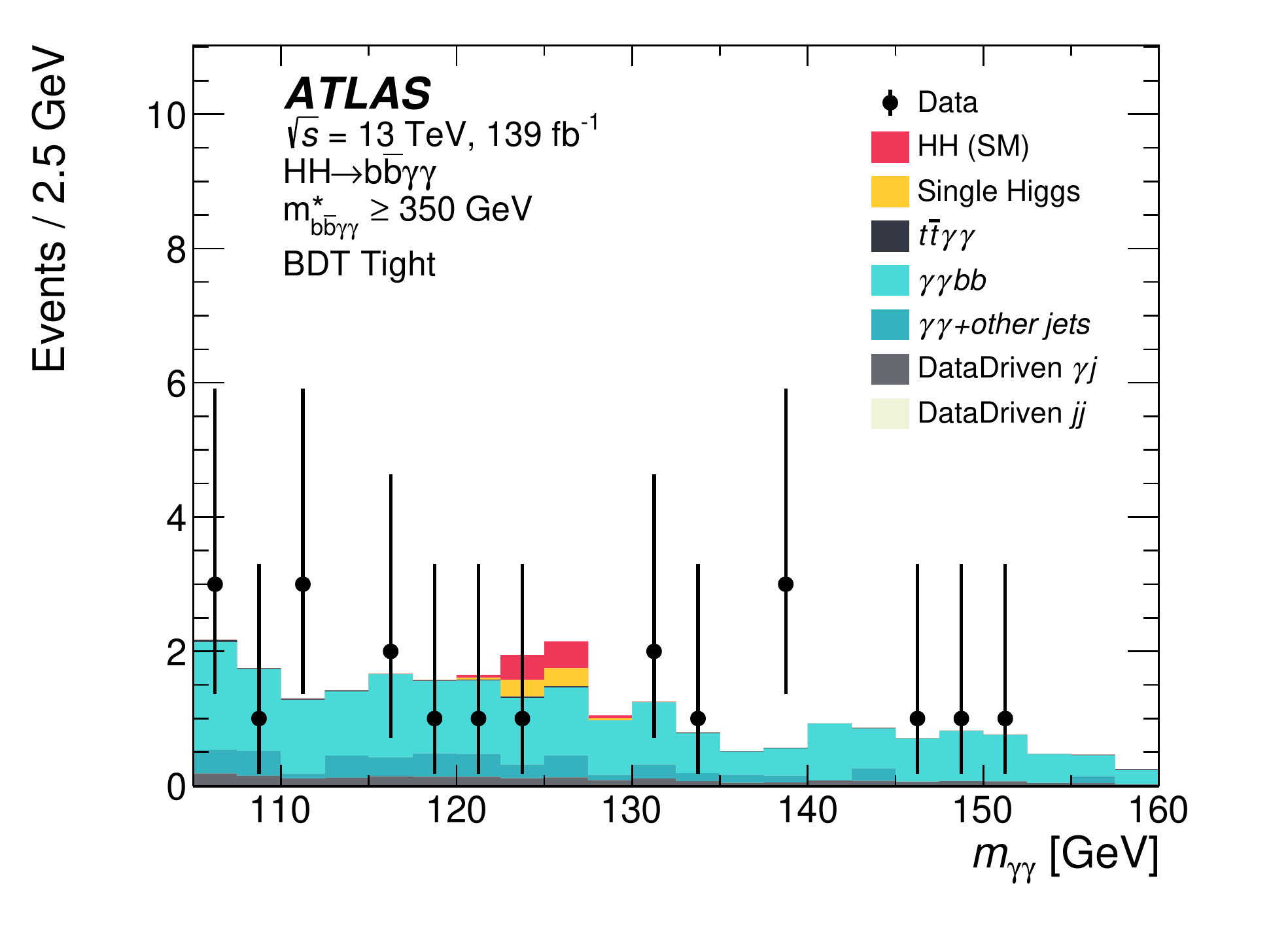}}
\subfloat[High mass BDT loose selection]{\includegraphics[width=0.4\textwidth]{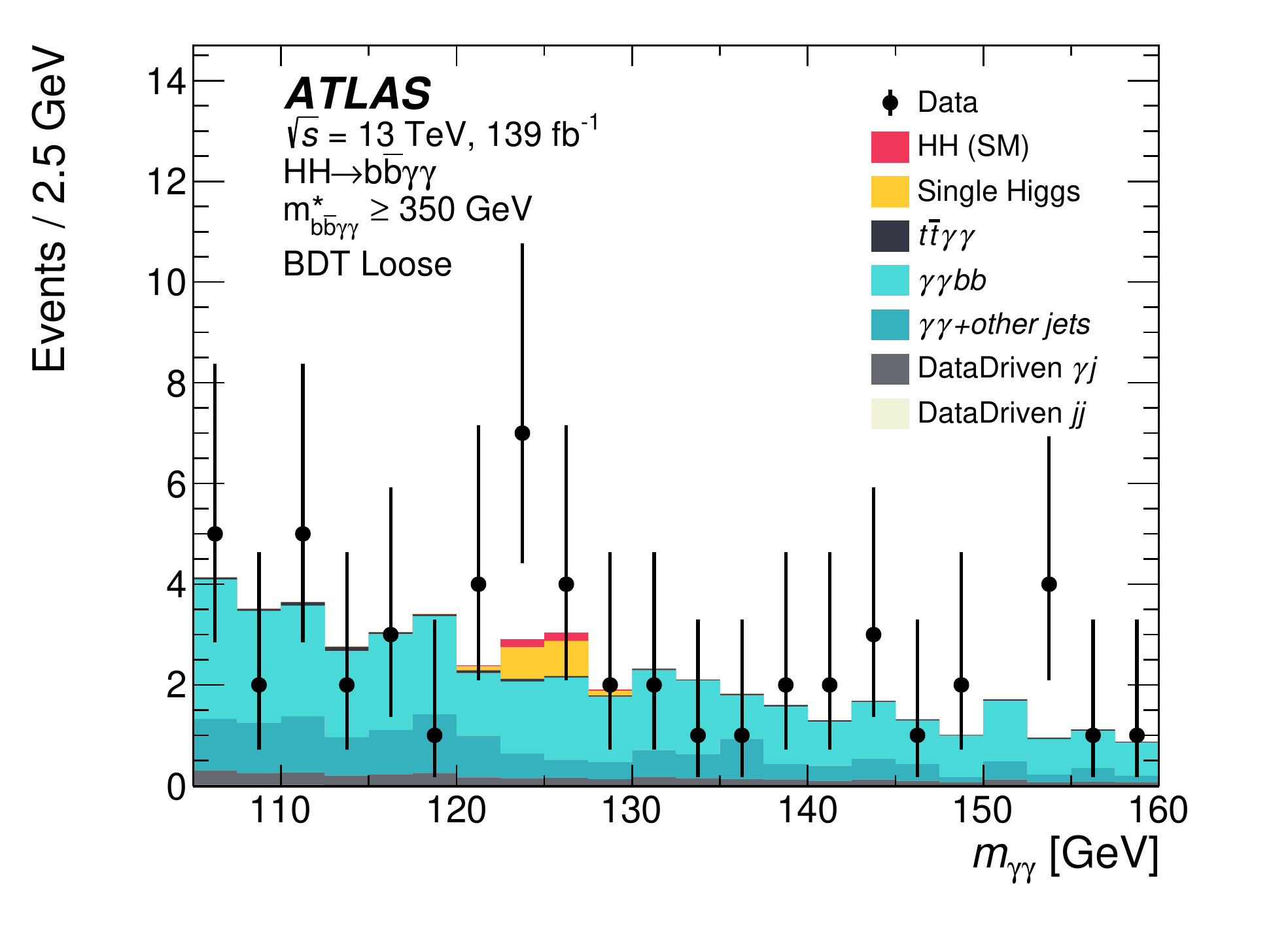}}\\
\subfloat[Low mass BDT tight selection]{ \includegraphics[width=0.4\textwidth]{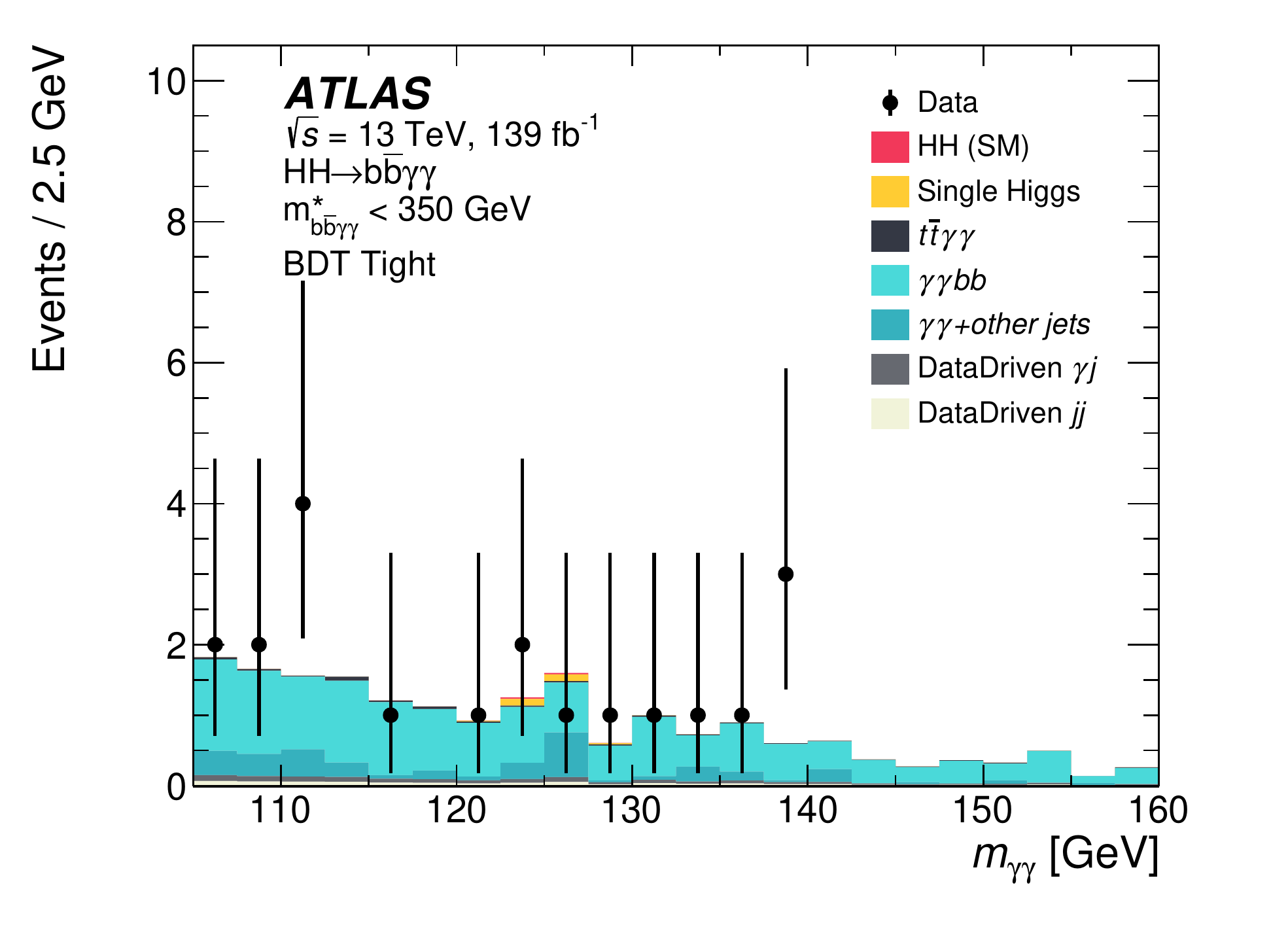}}
\subfloat[Low mass BDT loose selection]{\includegraphics[width=0.4\textwidth]{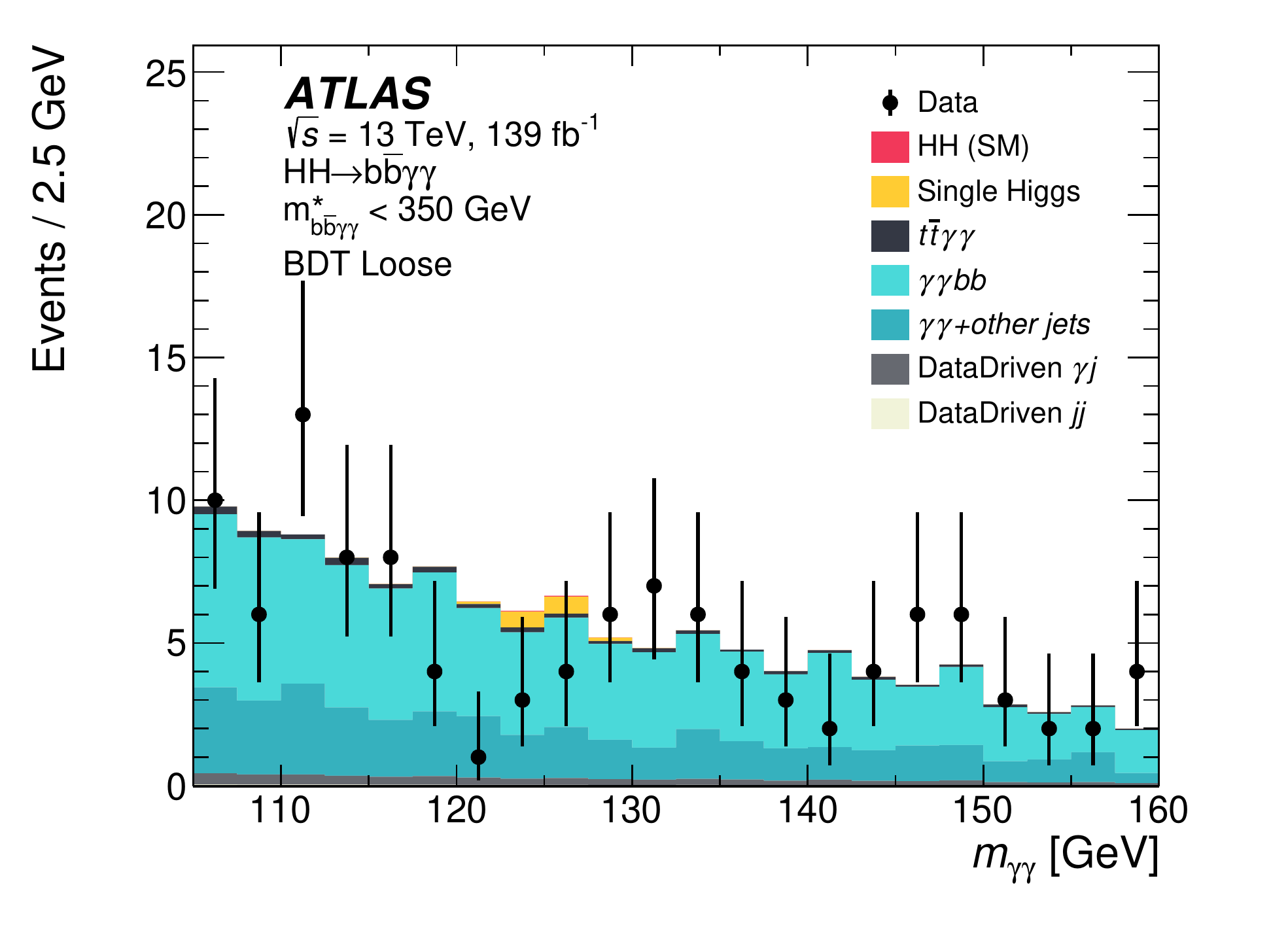}}
\caption{Distributions of \myy{} in all signal categories for the nonresonant $HH$ search: (a) high mass BDT tight, (b) high mass BDT loose, (c) low mass BDT tight, (d) low mass BDT loose.
The data-derived fractions of nonresonant $\gamma \gamma$, $\gamma$--jet or jet--$\gamma$, and dijet background are applied and the total background is normalized to the data sideband.}
\label{fig:dataMC_myy_nonRes}
\end{figure}
 
\begin{figure}
\centering
\subfloat[]{ \includegraphics[width=0.4\textwidth]{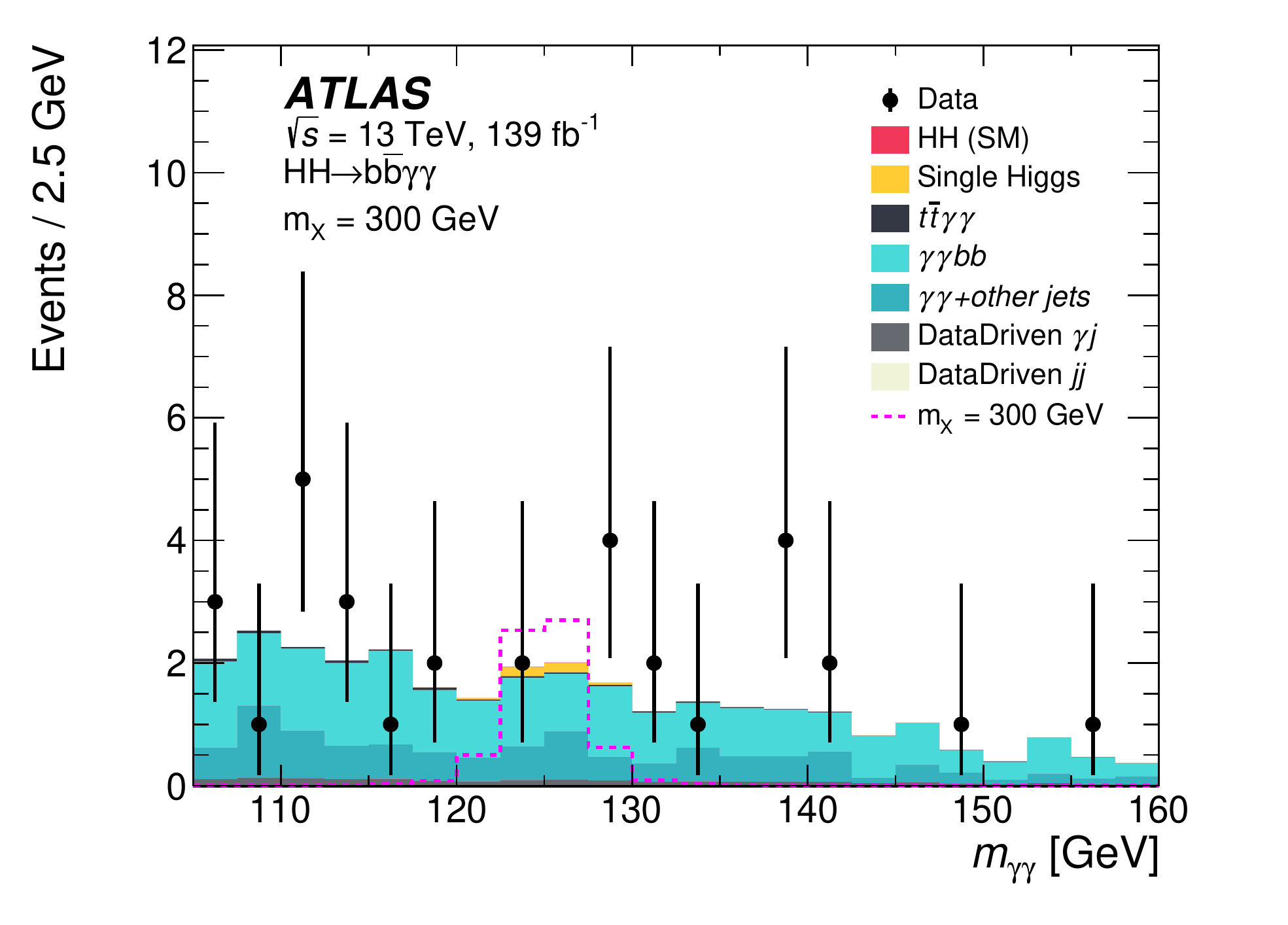}}
\subfloat[]{\includegraphics[width=0.4\textwidth]{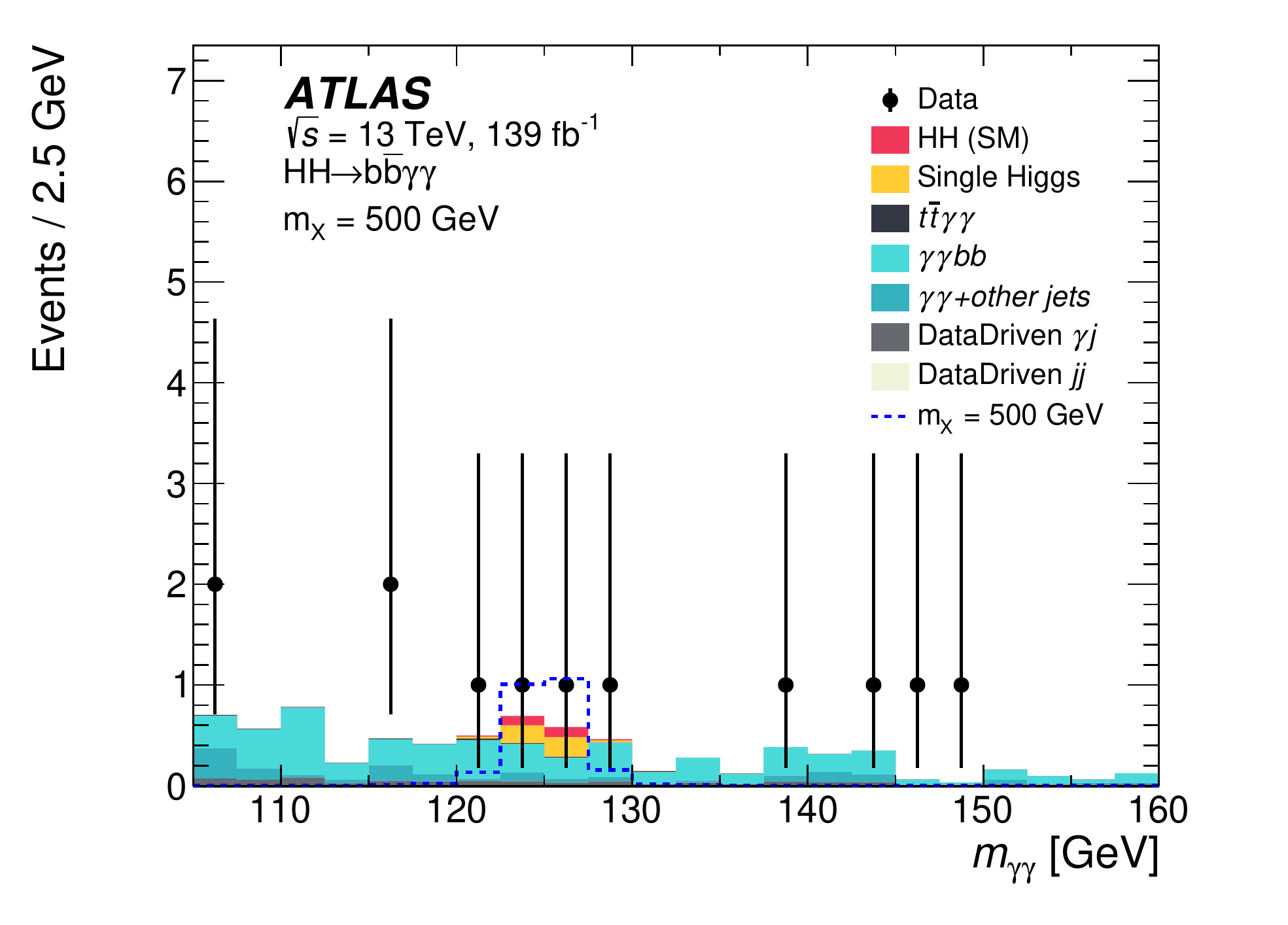}}
\caption{Distributions of \myy{} for the selections used for the resonance mass points (a) $m_X = 300$~\GeV{} and (b) $m_X = 500$~\GeV{} for the resonant search.
The data-derived fractions of nonresonant $\gamma \gamma$, $\gamma$--jet or jet--$\gamma$, and dijet background are applied and the total background is normalized to the data sideband.
The scalar resonance signal is scaled to a total production cross section $\sigma(pp\to X \to HH)=370$~fb  for $m_{X}=300~\GeV$ or $\sigma(pp\to X \to HH)=67$~fb for $m_{X}=500~\GeV$, corresponding to the expected exclusion limits presented in Section~\ref{sec:results_resonant}.}
\label{fig:dataMC_myy_Res}
\end{figure}

% End of text imported from the .//sections/object_event_selection.tex input file

% The next lines are included from the .//sections/signal_modelling.tex input file
\section{Signal and background parameterization}
\label{sec:sigbackpar}
 
The signal and backgrounds are extracted by fitting analytic functions to the diphoton invariant mass distribution in the range $105 < \myy < 160~\GeV$ in both the resonant and nonresonant $HH$ searches.
 
\subsection{$H \rightarrow \gamma \gamma$ parameterization}
For the di-Higgs signal and single-Higgs-boson background distributions, the parameterized forms are determined through fits to simulated samples and the expected normalizations are obtained from their theoretical cross sections multiplied by the product of the acceptance times efficiency from the simulation. The diphoton invariant mass distribution shapes are modeled with a double-sided Crystal Ball function~\cite{HIGG-2013-08,dscb1}, which is characterized by a Gaussian core and asymmetric power-law tails. This function allows the modeling of event distributions in which non-Gaussian tails can arise from experimental effects, such as photon energy mismeasurements. 
 
The shape parameters are determined by fitting the diphoton mass distribution in simulation for each category. The width of the fitted function is largely insensitive to the specific signal processes considered in the analysis, with maximum variations of approximately 10\%.
For the nonresonant search, the  parameterized form of \myy{} is obtained from the simulation of the \ggF\ and VBF $HH$ processes with $\kappa_{\lambda}=1$, described in Section~\ref{sec:simulation}. No significant dependence of the functional form on $\kappa_{\lambda}$ was found. 
For the resonant search, the functional form is obtained from the simulation of the heavy-resonance signals.
Table~\ref{tab:signalmyy1} shows the resolution parameter  of the  double-sided Crystal Ball functional form fit to the \myy{} distribution for simulated Higgs boson pair events in the nonresonant categories and for two different mass hypotheses for a heavy resonant signal. 
For both searches, the chosen functional forms are found to model both the single Higgs and di-Higgs boson events well.
As no statistically significant bias is observed in injection tests between the input and fitted signals, the same parameterized functions are used for both the single Higgs and  di-Higgs boson processes.
 
\begin{table}[htb!]
\centering
\caption{The resolution parameter of the  double-sided Crystal Ball functional form and the corresponding statistical uncertainty are obtained from the fit to the \myy{} distribution for simulated Higgs boson pair events for the nonresonant categories and for two different mass hypotheses for a heavy-resonance signal.
\label{tab:signalmyy1}}
\begin{tabular}{lc}
\toprule
\midrule
Category        &$\sigma_{\text{DSCB}}$~[\GeV]\\
\midrule
\midrule
High mass BDT tight        &$1.33 \pm 0.01$\\
High mass BDT loose        &$1.47 \pm 0.02$\\
Low mass BDT tight         &$1.50 \pm 0.06$\\
Low mass BDT loose         &$1.64 \pm 0.03$\\
\midrule
Resonant $m_X = 300$~\GeV  &$1.78 \pm 0.02$\\
Resonant $m_X = 500$~\GeV  &$1.46 \pm 0.01$\\
\bottomrule
\end{tabular}
\end{table}

\subsection{Background parameterization}
\label{sec:backpar}
The continuum diphoton background is modeled using a functional form chosen by fitting a highly populated MC background template.
Given the high $\gamma\gamma$ purity quoted in Section~\ref{sec:background_diphoton}, the background template is constructed in each category from the $\gamma\gamma$+jets simulation which is normalized to the data sideband in the mass windows of $105$--$120$~\GeV\ and $130$--$160$~\GeV\ in the $m_{\gamma\gamma}$ spectrum.
 
The potential bias associated with the choice of a specific analytic function to model the continuum background is estimated for each category as prescribed in Refs.~\cite{HIGG-2013-08,ATL-PHYS-PUB-2020-028}. It is obtained as the signal event yield extracted from a signal-plus-background fit to the background-only diphoton invariant mass distribution in the range $105 < m_{\gamma\gamma} < 160$~\GeV. This bias is also called the ``spurious signal''.
The number of fitted signal events is computed for Higgs boson masses in intervals of $1$~\GeV\ from $121$~\GeV\ to $129$~\GeV. The bias is taken as the largest number of fitted spurious signal events in this $8$~\GeV\ mass window.
Of the different analytic functions that are tested, the one with the smallest number of parameters is chosen from the functions with a spurious signal smaller than 20\% of the data's statistical uncertainty plus two times the MC statistical uncertainty.
In each category of the nonresonant search and in all the analysis regions defined in the resonant $HH$ search, an exponential function $\exp(a\cdot \myy)$ is found to be the best choice since it has the smallest number of degrees of freedom and yields a consistently small bias.
Wald tests~\cite{Wald:1945} on data show that the data do not prefer a higher-degree functional form to the exponential form. The bias is found to be at most 40\% (60\%) of the expected error on the fitted signal yield originating from the data statistics for the nonresonant (resonant) search. The difference in shape between the simulated events and the exponential form measured in the sidebands is found to have a negligible impact on the spurious signal. 
To study the impact on the spurious signal from the difference in shape between the MC background template and the data sideband, the MC template is reweighted to data using a linear function of $m_{\gamma\gamma}$ derived from the MC-data difference. The spurious signal value evaluated from the reweighted MC template is found to be compatible with the nominal one used as the systematic uncertainty.
% End of text imported from the .//sections/signal_modelling.tex input file
 
% The next lines are included from the .//sections/systematics.tex input file
\section{Systematic uncertainties}
\label{sec:systematics}
The sensitivity of the analysis is limited by the statistical precision. The assessment of the systematic uncertainties is described below and their impact on the results is discussed in Section~\ref{sec:syst_impact}. The uncertainty in the integrated luminosity of the full Run~2 data set is 1.7\%~\cite{ATLAS-CONF-2019-021}, obtained using the LUCID-2 detector~\cite{LUCID2} for the primary luminosity measurements.

The continuum background processes of the analysis are estimated from data and are subject to uncertainties related to the potential bias arising from the chosen background model, as described in Section~\ref{sec:backpar}. The background functional form bias is assessed as an additional uncertainty in the total number of signal events in each category.
For the single Higgs boson and di-Higgs boson production processes, both of which are estimated using simulation, experimental and theoretical systematic uncertainties are propagated through the full analysis procedure, as described in the following.
 
The efficiency of the diphoton trigger used to select events is evaluated in simulation and data using a bootstrap method and radiative $Z$ decays~\cite{TRIG-2018-05}. The difference between data and MC is taken as a systematic uncertainty.
In the diphoton invariant mass range of $105 < m_{\gamma\gamma} < 160$~\GeV, the trigger efficiency uncertainty affects the acceptance by 1\% in each category. The uncertainty in the vertex selection efficiency is assessed by comparing the efficiency of finding photon-pointing vertices in $Z \rightarrow e^+e^-$ events in data with that in simulation~\cite{HIGG-2016-08}. The resulting uncertainty is found to have a negligible effect on the signal selection efficiency.

The systematic uncertainties from the photon identification and isolation efficiencies are estimated following the prescriptions in Ref.~\cite{EGAM-2018-01}. They are evaluated by varying the correction factors for photon selection efficiencies in simulation by the corresponding uncertainties and affect the diphoton selection efficiency.
The experimental uncertainties in the photon energy scale and resolution are obtained from Ref.~\cite{EGAM-2018-01}.
 
The jet energy scale and resolution uncertainties~\cite{JETM-2018-05} affect the $\mbb$ distribution, while flavor-tagging uncertainties affect the acceptance of the analysis categories. The experimental uncertainties in jet energy scale and resolution are propagated to the \etmiss calculation. In addition, the uncertainties in the scale and resolution of the \etmiss\ soft term are evaluated by using the method described in Ref.~\cite{PERF-2016-07}.
The flavor-tagging uncertainties for $b$- and $c$-jets are estimated in \ttbar{} events~\cite{FTAG-2018-01,ATLAS-CONF-2018-001}, while the misidentification uncertainty of light-flavor jets is determined using dijet events~\cite{ATLAS-CONF-2018-006}.
Additional uncertainties from the $b$-jet momentum correction accounting for the presence of muons and neutrinos are found to be negligible.
 
For single Higgs boson and SM $HH$ production, the effects of theoretical scale uncertainties due to missing higher-order corrections on the production rates are estimated by varying the factorization and renormalization scales up and down from their nominal values by a factor of two, recalculating the cross section in each case, and taking the largest deviation from the nominal cross section as the uncertainty. The uncertainties in SM $HH$ ggF production arising from the choice of renormalization scheme and scale of the top-quark mass and their combination with those from factorization and renormalization scale variations are based on Ref.~\cite{Baglio:2020wgt}. The uncertainties in the cross sections, including effects of PDF+\alphas uncertainties, and the uncertainties in the $H \rightarrow \yy $ and $H \rightarrow \bb$ branching fractions, are taken from Refs.~\cite{DiMicco:2019ngk,deFlorian:2016spz}. The uncertainty in the value of the Higgs boson mass is considered~\cite{HIGG-2014-14}. An additional 100\% uncertainty is assigned to
the single-Higgs-boson ggF and  VBF production modes and to Higgs boson production in association with a $W$ boson. This is motivated by studies of heavy-flavor production in association with top-quark pairs~\cite{TOPQ-2017-12,TOPQ-2012-16-corr} and of $W$ boson production in association with $b$-jets~\cite{STDM-2012-11}.
No additional heavy-flavor uncertainty is assigned to the single-Higgs-boson \tth{} and $ZH$ production modes, where the dominant heavy-flavor production is already accounted for in the LO process.
For all samples, the uncertainty related to the choice of parton showering model is evaluated by comparing the predictions of the nominal MC samples and alternative samples using a different parton showering model.
In addition, for the nonresonant $HH$ production processes, a systematic uncertainty is assigned to the $\kappa_{\lambda}$ reweighting, as discussed in Section~\ref{sec:simulation}.

In the resonant search, the systematic uncertainty sources considered are the same as for the nonresonant case, except that uncertainties due to the finite order of the QCD calculations are not assigned to the resonant signal cross section.
In this search the SM $HH$ production processes are considered as background. 
% End of text imported from the .//sections/systematics.tex input file
 
% The next lines are included from the .//sections/Results.tex input file
\section{Results}
The statistical framework used to derive the results for both the nonresonant and resonant searches is described in the following.
 
\subsection{Statistical framework}
For both the nonresonant and resonant searches, the results of the analysis are obtained from a maximum-likelihood fit of the \myy distribution in the range $105 < \myy < 160~\GeV$, performed simultaneously over all relevant categories described in Section~\ref{sec:event_selection}. The likelihood function is defined in Eq.~(\ref{eqn:likelihood}):
 
\begin{equation}
\label{eqn:likelihood}
\mathcal{L} = {\displaystyle\prod_{c}\left(\mathrm{Pois}(n_c|N_c(\pmb{\theta}))\cdot{\displaystyle\prod_{i=1}^{n_c} f_c(m_{\gamma\gamma}^i,\pmb{\theta})} \cdot G( \pmb{\theta})\right)},
\end{equation}
 
where for each event $i$ in a category $c$, $n_c$ is the observed number of events, $N_c$ is the expected number of events, $f_c$ is the value of the probability density function (pdf), $\pmb{\theta}$ are nuisance parameters, and $G(\pmb{\theta})$ are constraint pdfs for the nuisance parameters.
 
The expected number of events $N_c$, defined in Eq.~(\ref{eqn:expected_number_events}), is the sum of the expected yields from di-Higgs boson production processes, single Higgs boson production, the nonresonant background, as well as the spurious-signal uncertainty:
 
\begin{equation}
\label{eqn:expected_number_events}
N_c( \pmb{\theta}) = \mu \cdot N_{HH,\mathrm{c}}( \pmb{\theta}_{HH}^\mathrm{yield}) + N^{\mathrm{res}}_{\mathrm{bkg,c}}( \pmb{\theta}_{\mathrm{res}}^\mathrm{yield}) + N_{\mathrm{SS,c}} \cdot  \pmb{\theta}^{\mathrm{SS, c}} + N^{\mathrm{nonres}}_{\mathrm{bkg,c}},
\end{equation}
 
where $\mu$ is the signal strength, $ \pmb{\theta}^\mathrm{SS, c}$ represent the nuisance parameters associated with the background function bias and $ \pmb{\theta}^\mathrm{yield}$ represent the nuisance parameters affecting the event yield, as detailed in Section~\ref{sec:systematics}. Correlation of the nuisance parameters across different signal and background components, as well as categories, is taken into account. In the case of the nonresonant search, $N^{\mathrm{res}}_{\mathrm{bkg,c}} =  N_{H,\mathrm{c}}$, while in the case of the resonant analysis $N^{\mathrm{res}}_{\mathrm{bkg,c}} =  N_{H,\mathrm{c}} +  N_{\mathrm{SM} HH,\mathrm{c}}$.
 
The probability density function $f_c$ represents the shape information. The sum of the double-sided Crystal Ball functions modeling the $HH$ production processes, single Higgs boson production, and the spurious signal, and of the analytic function modeling the nonresonant background as described in  Section~\ref{sec:backpar}, is shown in Eq.~(\ref{eqn:pdf}):
\begin{multline}
\label{eqn:pdf}
f_{c}(\myy, \pmb{\theta}) = [\mu \cdot N_{HH,\mathrm{c}}( \pmb{\theta}_{HH}^\mathrm{yield}) \cdot f_{HH,\mathrm{c}}(\myy,  \pmb{\theta}_{HH}^\mathrm{shape}) + N^{\mathrm{res}}_{\mathrm{bkg, c}}( \pmb{\theta}_{\mathrm{res}}^\mathrm{yield}) \cdot f^{\mathrm{res}}_\mathrm{bkg, c}(\myy,  \pmb{\theta}_{\mathrm{res}}^\mathrm{shape}) \\ + N_\mathrm{SS,c} \cdot  \pmb{\theta}_{HH}^\mathrm{SS, c} \cdot f_{HH,\mathrm{c}}(\myy,  \pmb{\theta}_{HH}^\mathrm{shape}) + N^{\mathrm{nonres}}_{\mathrm{bkg,c}} \cdot f^{\mathrm{nonres}}_{\mathrm{bkg,c}}(\myy,  \pmb{\theta}_{\mathrm{nonres}}^\mathrm{shape})]/N_c( \pmb{\theta}_{\mathrm{nonres}}^\mathrm{yield}),
\end{multline}
 
where $ \pmb{\theta}^\mathrm{shape}$ represent nuisance parameters related to the shape variations of the functional forms. When a nuisance parameter is related to shape and yield variations at the same time, the two effects are correlated.
 
The nominal yields of the resonant background processes are initially set to values from simulation.
The likelihood function includes all the nuisance parameters that describe the systematic uncertainties. The signal strength is a free parameter in the fit.
The measurement of the parameter of interest is carried out using a statistical test based on the profile likelihood ratio~\cite{Cowan:2010js}, as shown in Eq.~(\ref{eq:Lambda}):
\begin{equation}
\Lambda(\mu) =
\frac{\mathcal{L}(\mu, \hat{\hat{\pmb{\theta}}}(\mu)) }{\mathcal{L}(\hat{\mu}, \hat{\pmb{\theta}})} \;,
\label{eq:Lambda}
\end{equation}
where $\mu$ and $\pmb{\theta}$ are respectively the parameter of interest and the nuisance parameters.
In the numerator, the nuisance parameters are set to their profiled values $\hat{\hat{\pmb{\theta}}}(\mu)$, which maximize the likelihood function for a fixed value of the parameter of interest $\mu$. In the denominator, the parameter of interest and the nuisance parameters are set to the values $\hat{\mu}$ and $\hat{\pmb{\theta}}$, respectively, which jointly maximize the likelihood.
 
In the absence of signal, exclusion limits are set on Higgs boson pair production in the \yybb{} final state. The limits for both nonresonant
and resonant production are calculated using the $\textrm{CL}_{\textrm{S}}$ method~\cite{Read:2002hq}, with the profile-likelihood-ratio-based test
statistic $\tilde{q}_{\mu}$, defined in Eq.~(\ref{eqn:plr})~\cite{Cowan:2010js}:
 
\begin{equation}
\label{eqn:plr}
\tilde{q}_\mu = \begin{cases}
-2\ln\frac{\Lambda(\mu,\hat{\hat{\pmb{\theta}}}(\mu))}{\Lambda(0,\hat{\hat{\pmb{\theta}}}(0))} & \hat{\mu} < 0,\\
-2\ln\frac{\Lambda(\mu,\hat{\hat{\pmb{\theta}}}(\mu))}{\Lambda(\hat{\mu},\hat{\pmb{\theta}}(\mu))} & 0 \leq \hat{\mu} \leq \mu,\\
0 & \hat{\mu} > \mu.
\end{cases}
\end{equation}
 
The asymptotic approximation~\cite{Cowan:2010js} is used for the test-statistic distribution. The inaccuracy of the asymptotic approximation is checked with pseudo-experiment studies for the key results reported in this paper.

\subsection{Nonresonant search results}

Figure~\ref{fig:nonres_sideband_fit} shows the background fits to the data. No significant excess over the SM background expectations is found, as summarized in Table~\ref{tab:yields1}.
The statistical analysis sets a  95\% CL upper limit on the nonresonant $HH$ production cross section at 130~fb, while 180~fb is expected.  An observed (expected) upper limit  at 95\% CL on the signal strength of 4.2 (5.7) times the SM prediction is obtained.  The  expected constraints are obtained for a background hypothesis excluding $pp \rightarrow HH$ production.
For the upper limits on the cross section, all theoretical uncertainties are included, except those related to the signal cross section itself, while constraints on the signal strength are computed including uncertainties in the predicted signal cross section. A check of the results that quantifies the upper limits by using pseudo-experiments is performed and the increase of the limit value relative to the asymptotic approximation is found to be less than 8\% for both the  observed and expected upper limits. A check of the expected upper limits using pre-fit values for the nuisance parameters associated with the systematic uncertainties is performed and an increase of 4\% relative to the nominal result is found. The difference is dominated by the contribution from the spurious signal.
 
Upper limits on the $HH$ production cross section are also computed as a function of $\klambda$, as shown in  Figure~\ref{fig:nonres_expected_klambda}. For this purpose, single-Higgs-boson production cross sections and Higgs boson decay branching ratios are assumed to have SM values, and the coupling strength between the Higgs boson and other particles are also set to their SM values~\cite{HIGG-2018-57}.
The theory uncertainties related to the signal cross section are not included.
 
The expected constraints on $\klambda$ at 95\% CL, as obtained with a background hypothesis excluding $pp \to HH$ production, are $[-2.4, 7.7]$, whereas the observed constraints are $[-1.5, 6.7]$  at 95\% CL.
The inclusion of the VBF $HH$ production mode tightens the constraints by about 5\% relative to an alternative fit considering only the ggF production mode.

\begin{table*}[htb]
\footnotesize
\caption{
The number of data events observed in the  $120<\myy<130~\GeV$ window,
the number of $HH$ signal events expected for $\klambda=1$ and for $\klambda=10$,
and events expected for single Higgs boson production (estimated using MC simulation),
as well as for continuum background. For the single Higgs boson, ``Rest'' includes the following production modes: VBF, $WH$, $tHq$, and $tHW$.
The values are obtained from a fit of the Asimov data set~\cite{Cowan:2010js} generated under the SM signal-plus-background hypothesis, $\klambda = 1$. The continuum background component  of the Asimov data set is obtained from the fit of the data sideband.
The uncertainties in $HH$ signals and
single Higgs boson background include the systematic uncertainties discussed in Section~\ref{sec:systematics}.
The uncertainty in the continuum background is given by the sum in quadrature of the statistical uncertainty
from the fit to the data and the spurious-signal uncertainty.
}
\begin{center}
\begin{tabular}{lcccc}
\toprule
\toprule
&   High mass & High mass &  Low mass  & Low mass \\
&   BDT tight & BDT loose &  BDT tight & BDT loose \\
\midrule
Continuum background                & $4.9^{+1.1}_{-1.3}$      & $9.5^{+1.5}_{-1.7}$   & $3.7^{+0.9}_{-1.1}$      & $24.9^{+2.3}_{-2.5}$      \\[4pt]
Single Higgs boson background       & $0.67^{+0.29}_{-0.13}$   & $1.6^{+0.6}_{-0.2}$   & $0.23^{+0.09}_{-0.03}$    & $1.40^{+0.33}_{-0.16}$      \\[4pt]
\phantom{Single Higgs}\ggF$+bbH$    & $0.26^{+0.28}_{-0.16}$   & $0.4^{+0.5}_{-0.2}$   & $0.07^{+0.08}_{-0.04}$    & $0.27^{+0.27}_{-0.16}$    \\[4pt]
\phantom{Single Higgs}\ttH          & $0.19^{+0.03}_{-0.03}$   & $0.49^{+0.09}_{-0.07}$ & $0.107^{+0.022}_{-0.017}$ & $0.75^{+0.13}_{-0.11}$ \\[4pt]
\phantom{Single Higgs}$ZH$          & $0.142^{+0.035}_{-0.025}$ & $0.48^{+0.09}_{-0.07}$ & $0.040^{+0.020}_{-0.014}$ & $0.27^{+0.06}_{-0.04}$ \\[4pt]
\phantom{Single Higgs}Rest          & $0.074^{+0.032}_{-0.014}$ & $0.16^{+0.07}_{-0.03}$ & $0.012^{+0.008}_{-0.004}$ & $0.111^{+0.030}_{-0.012}$ \\[4pt]
\midrule
SM $HH (\klambda=1)$ signal         & $0.87^{+0.10}_{-0.18}$   & $0.37^{+0.04}_{-0.07}$  & $0.049^{+0.006}_{-0.010}$ & $0.078^{+0.008}_{-0.015}$   \\[4pt]
\phantom{Single Higgs}\ggF          & $0.86^{+0.10}_{-0.18}$   & $0.35^{+0.04}_{-0.07}$  & $0.046^{+0.006}_{-0.010}$ & $0.072^{+0.008}_{-0.015}$  \\[4pt]
\phantom{Single Higgs}VBF           & $(12.6^{+1.3}_{-1.2})\cdot 10^{-3}$ & $(16.1^{+1.4}_{-1.2})\cdot 10^{-3}$ & $(3.2^{+0.4}_{-0.4})\cdot 10^{-3}$ & $(6.9^{+0.5}_{-0.6})\cdot 10^{-3}$ \\[4pt]
\midrule
Alternative $HH (\klambda=10)$ signal  &  $6.5^{+1.0}_{-0.8}$ & $3.6^{+0.6}_{-0.4}$   & $4.5^{+0.7}_{-0.6}$      & $8.5^{+1.3}_{-1.0}$   \\
\midrule
Data                                &  2  & 17 & 5  & 14 \\
\bottomrule
\bottomrule
\end{tabular}
\label{tab:yields1}
\end{center}
\end{table*}

\begin{figure}[htbp]
\centering
\subfloat[High mass BDT tight]{\includegraphics[width=0.45\textwidth]{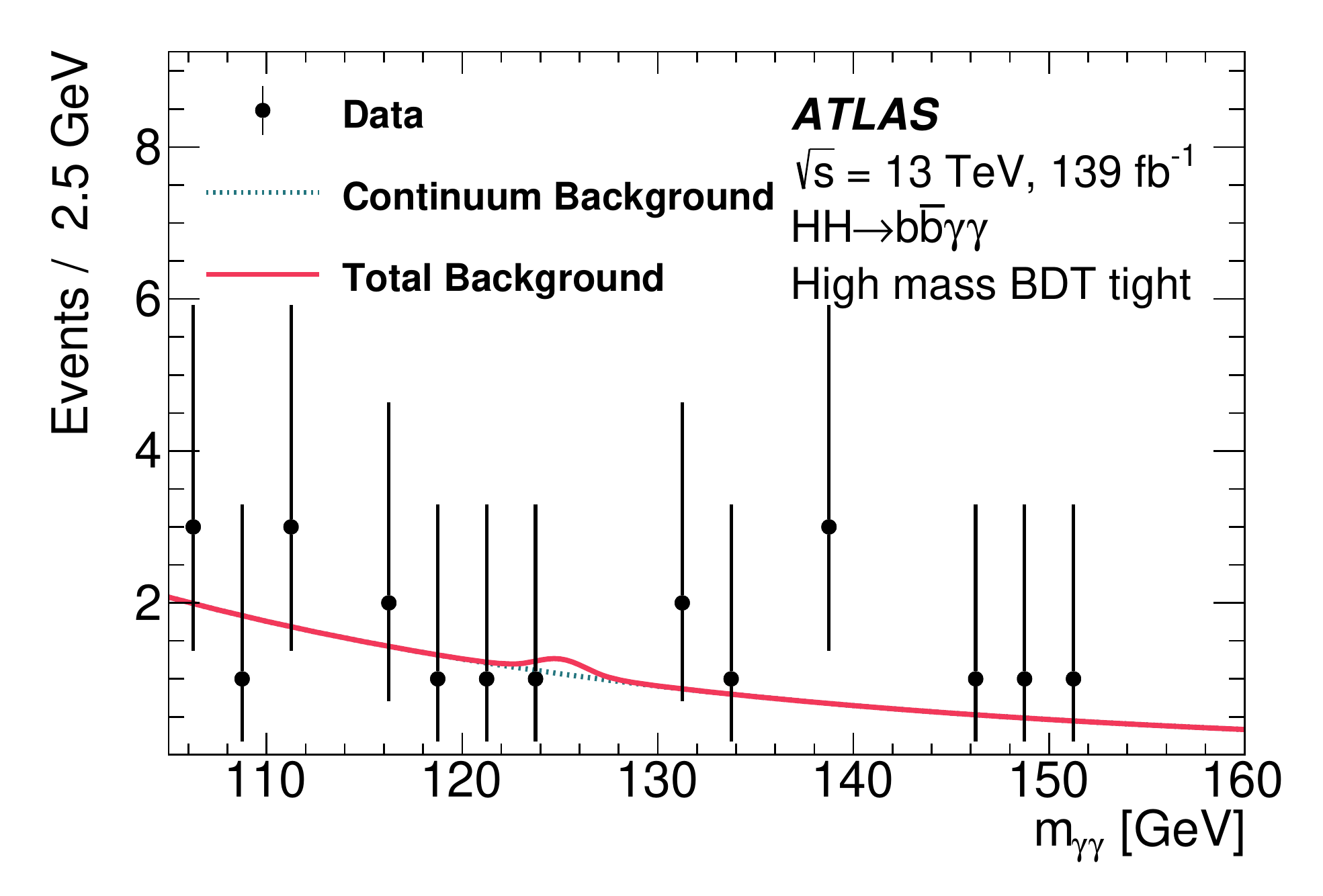}}
\subfloat[High mass BDT loose]{\includegraphics[width=0.45\textwidth]{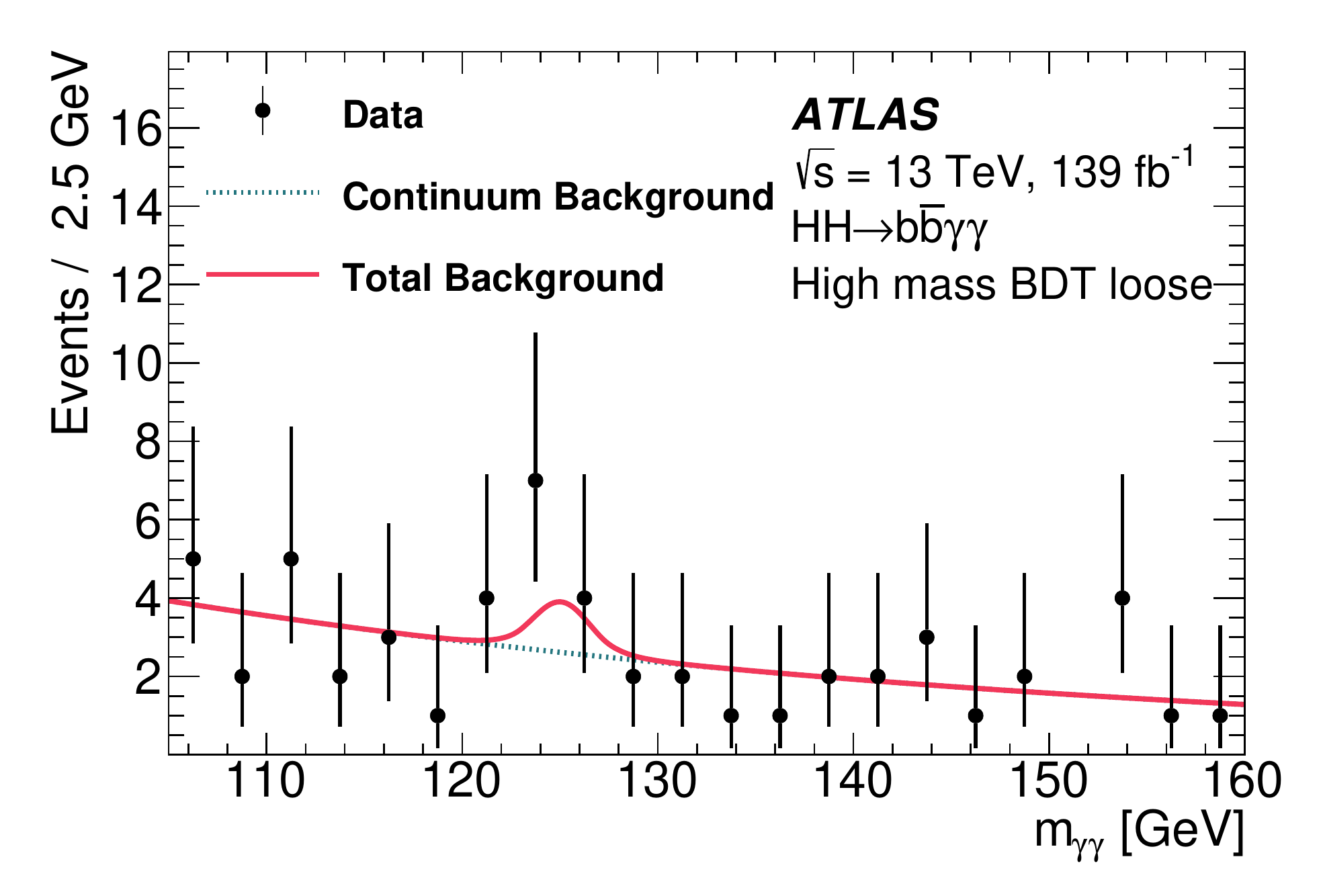}}\\
\subfloat[Low mass BDT tight]{\includegraphics[width=0.45\textwidth]{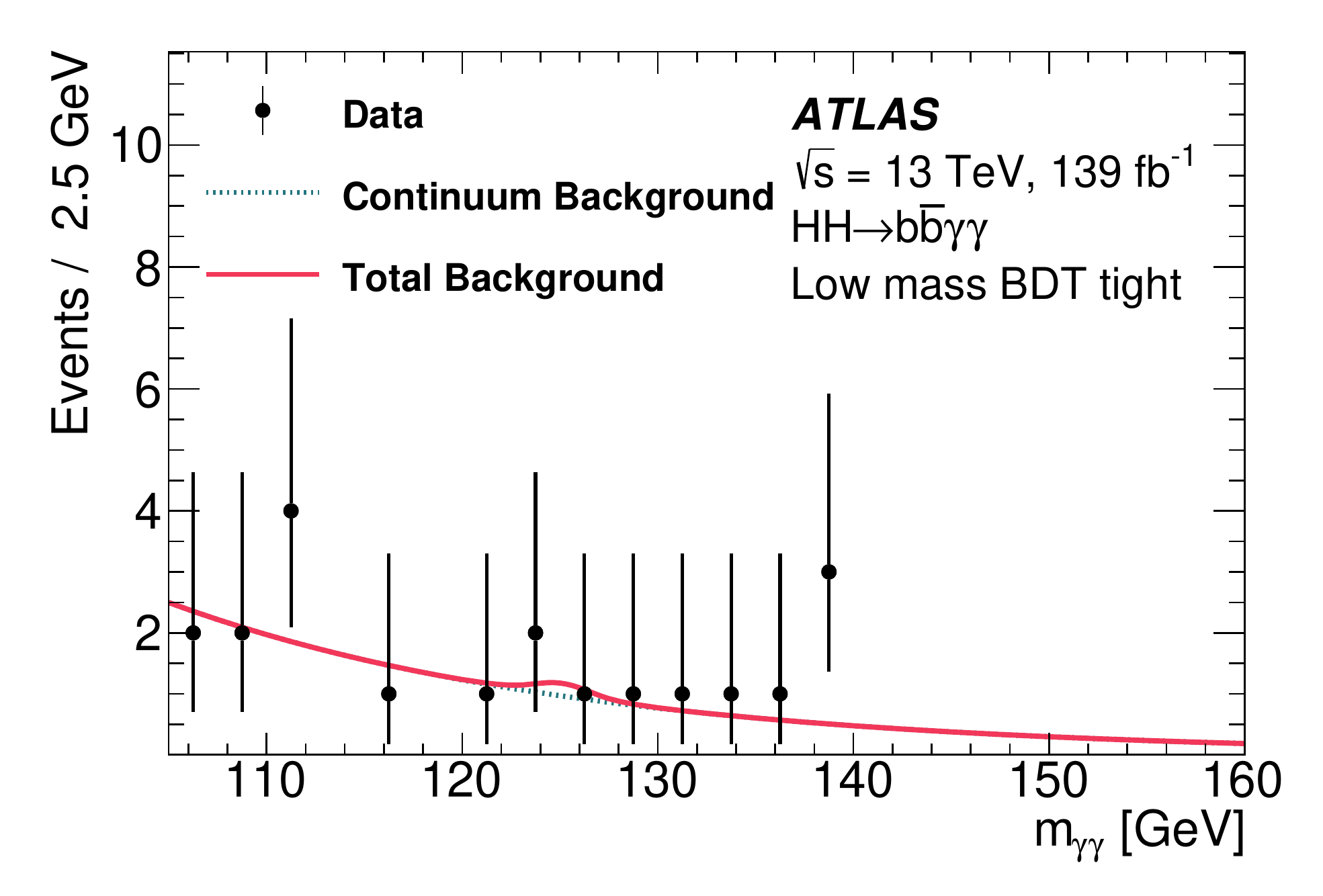}}
\subfloat[Low mass BDT loose]{\includegraphics[width=0.45\textwidth]{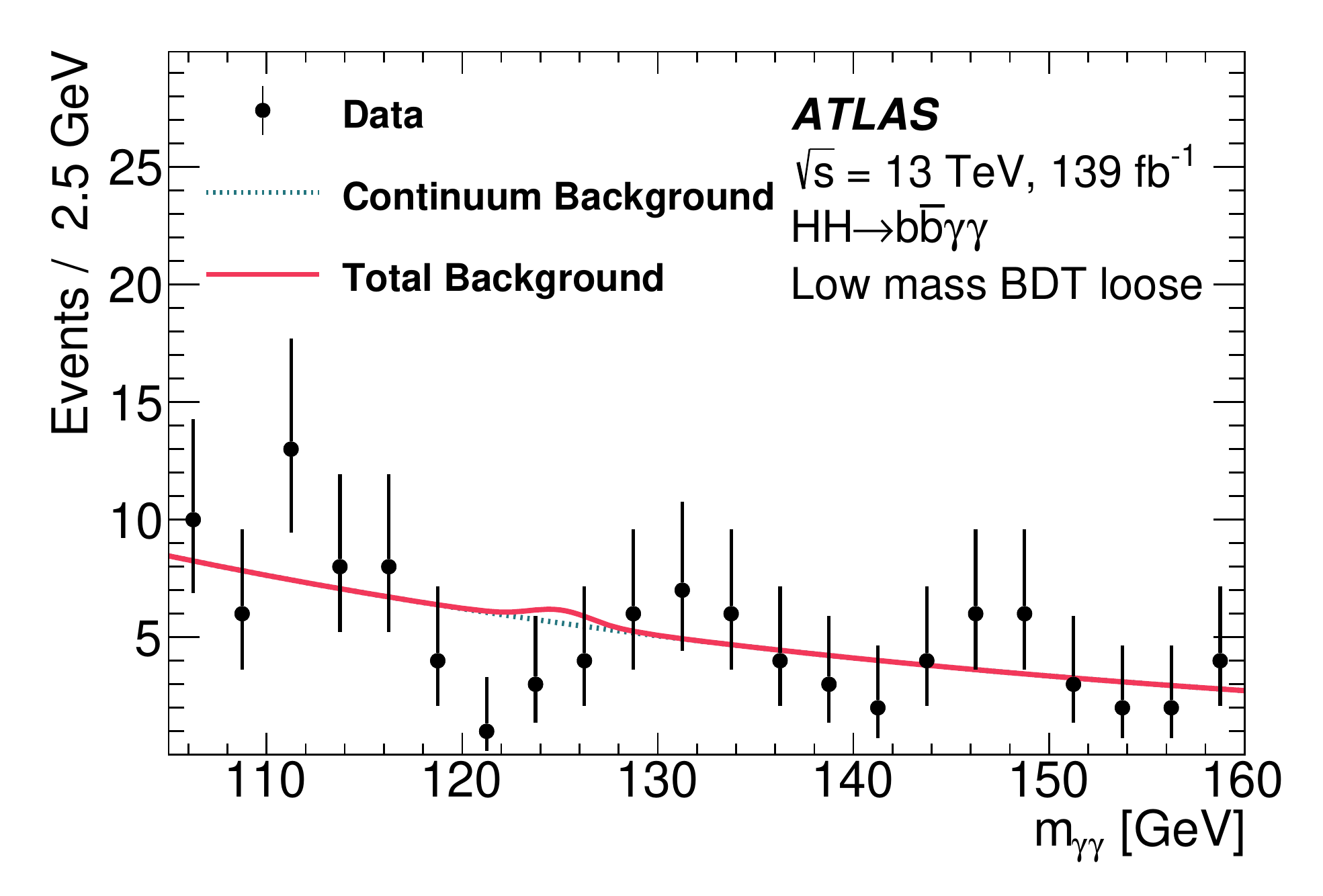}}
 
\caption{Data are compared with the background-only fit for the four categories of the nonresonant search: (a) high mass BDT tight, (b) high mass BDT loose, (c) low mass BDT tight, and (d) low mass BDT loose. Both the continuum background and the background from single Higgs boson production are considered. The data points in the Figure are the same as shown in Figure~\ref{fig:dataMC_myy_nonRes}.}
\label{fig:nonres_sideband_fit}
\end{figure}

\begin{figure}[htbp]
\centering
\includegraphics[width=0.65\textwidth]{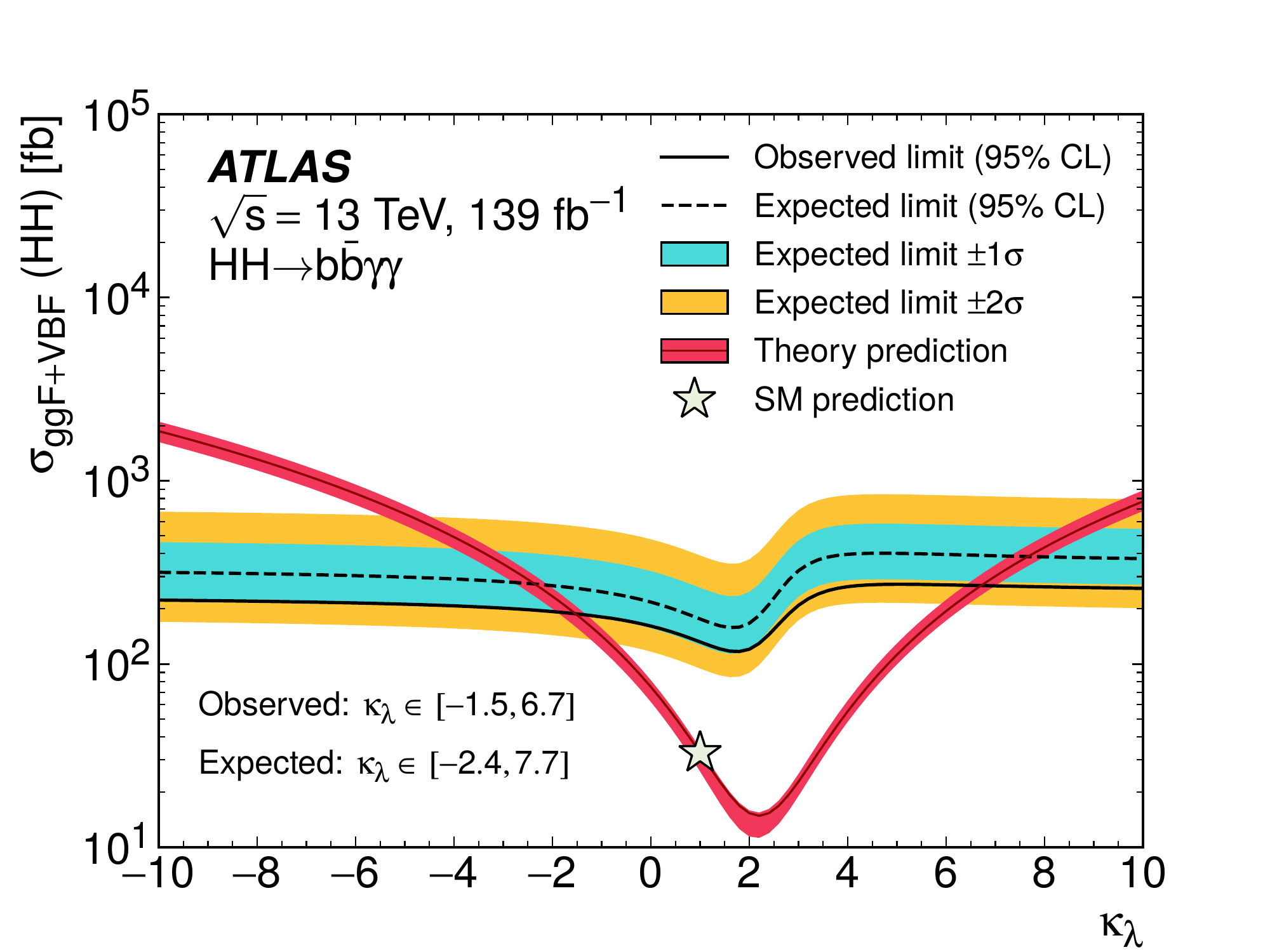}
\caption{Observed and expected limits at 95\% CL on the cross section of nonresonant Higgs boson pair production as a function of the Higgs boson self-coupling modifier $\kappa_{\lambda}= \lambda_{HHH}/\lambda^{\textrm{SM}}_{HHH}$. The  expected constraints on \klambda are obtained with a background hypothesis excluding $pp \rightarrow HH$ production.  The $\pm 1\sigma$ and $\pm 2\sigma$  variations about the expected limit due to statistical and systematic uncertainties are also shown. The theory prediction curve represents the scenario where all parameters and couplings are set to their SM values except for  $\kappa_{\lambda}$. The uncertainty band of the  theory prediction curve shows the cross-section uncertainty.}
\label{fig:nonres_expected_klambda}
\end{figure}

An alternative statistical analysis consists in determining the best-fit value of the $\klambda$  coupling modifier. The best-fit value of $\klambda$ and its uncertainty are obtained by means of a negative log-likelihood scan. 
The coupling strengths of the Higgs boson to fermions and gauge bosons are set to their SM values. The values of the negative log-likelihood ratio, $-2 \ln \Lambda(\mu)$, as a function of $\klambda$ are shown in Figure~\ref{fig:kappa_lambda_measurement_obs}.
The Asimov data set~\cite{Cowan:2010js} is generated under the SM signal-plus-background hypothesis, $\klambda=1$.
All systematic uncertainties, including those of the theoretical prediction of the $HH$ production cross section, are included.  The best-fit value corresponds to $\klambda = 2.8^{+2.0}_{-2.2}$ ($^{+3.8}_{-4.3}$)  for the  1$\sigma$ (2$\sigma$) confidence interval. The expected value corresponds to $\klambda = 1.0^{+5.5}_{-2.4}$ ($^{+7.3}_{-4.2}$) 
for the  1$\sigma$ (2$\sigma$) confidence interval.
The second minimum in the expected likelihood scan curve corresponds to a similar fitted signal yield with respect to the \klambda point at the first minimum, which is a consequence of a higher cross section, but lower acceptance and worse signal-to-background separation. The $m_{HH}$ distribution has a different shape at each of the two minima, as shown in Figure~\ref{fig:mx-vs-kl}.
 
\begin{figure}[htbp]
\centering
\includegraphics[width=0.69\textwidth]{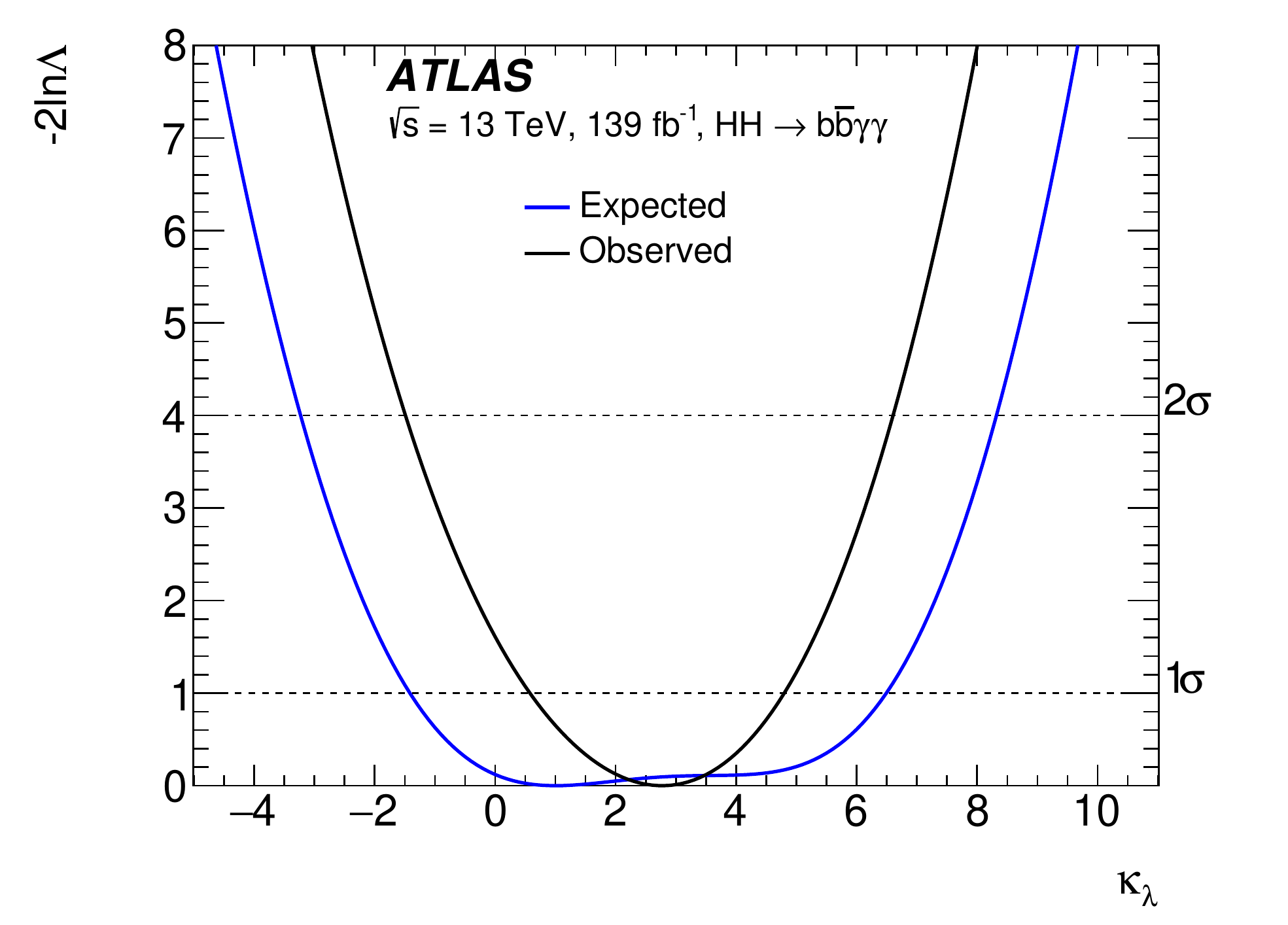}
\caption{Values of the negative log-profile-likelihood ratio ($-2 \ln \Lambda$) as a function of \klambda evaluated for the combination of all the categories of the nonresonant search. The coupling of the Higgs boson to fermions and gauge bosons is set to SM values in the profile likelihood calculation.
The expected result corresponds to a Asimov data set~\cite{Cowan:2010js} generated under the SM signal-plus-background hypothesis, $\klambda = 1$. All systematic uncertainties, including the theoretical uncertainties in the di-Higgs boson production cross section, are included.
The intersections of the solid curves and the horizontal dashed lines indicate the 1$\sigma$ and 2$\sigma$ confidence-level intervals. 
} \label{fig:kappa_lambda_measurement_obs}
\end{figure}

\FloatBarrier
\subsection{Resonant search results}
\label{sec:results_resonant}
 
Figure~\ref{fig:res_sideband_fit} shows the fit to the data of the resonant search for two benchmark values of the mass $m_{X}$ of a hypothetical scalar particle. No significant excess over the SM background expectations is found, as shown in Table~\ref{tab:yields2}. Figure~\ref{fig:res_expected_limit} shows the observed and expected upper limits at 95\% CL on the production cross section of a narrow-width scalar resonance. The observed (expected) upper limits vary between 640--44~fb (391--46~fb) in the range $251 \leq m_{X} \leq 1000~\GeV$.
A check on the upper limits using pseudo-expeirments is performed. For the expected limits, the results based on the pseudo-experiments are found to be up to 10\% higher compared with those derived based on the asymptotic approximation. As for the observed limits, the pseudo-experiments yield typically 10\% higher results compared with the asymptotic approximation in most of the $m_X$ range explored, and the difference increases to 15\% for the $m_X=700~\GeV$ signal hypothesis.

\begin{table*}[htb]
\small
\caption{The number of events observed in the  $120<\myy<130~\GeV$ window in data,
the number of events expected for scalar resonance signals of masses  $m_{X}=300~\GeV$ and $m_{X}=500~\GeV$
assuming a total production cross section $\sigma(pp\to X \to HH)$ equal to the observed exclusion limits of Figure~\ref{fig:res_expected_limit},
and events expected for SM $HH$ and single Higgs boson production (estimated using MC simulation),
as well as for continuum background.
The values are obtained from a fit of the Asimov data set~\cite{Cowan:2010js} generated under the signal-plus-background hypothesis. The continuum background component of the Asimov data set is obtained from the fit of the data sideband.
The uncertainties in the resonant signals and the SM $HH$ and
single-Higgs-boson backgrounds include the systematic uncertainties discussed in Section~\ref{sec:systematics}.
The uncertainty in the continuum background is given by the sum in quadrature of the statistical uncertainty
from the fit to the data and the spurious-signal uncertainty.
}
\begin{center}
\begin{tabular}{lcc}
\toprule
\toprule
&  $m_{X}=300~\GeV$ & $m_{X}=500~\GeV$ \\
\midrule
Continuum background             & $5.5^{+1.3}_{-1.5}$       & $1.6^{+0.6}_{-0.9}$  \\[4pt]
Single Higgs boson background    & $0.34^{+0.14}_{-0.07}$     & $0.40^{+0.18}_{-0.08}$  \\[4pt]
SM $HH$ background               & $0.021^{+0.005}_{-0.009}$  & $0.20^{+0.09}_{-0.09}$  \\[4pt]
\midrule
$X \to HH$ signal                & $6.1^{+0.9}_{-0.8}$       & $6.1^{+0.8}_{-0.6}$  \\
\midrule
Data                             & 6  & 4  \\
\bottomrule
\bottomrule
\end{tabular}
\label{tab:yields2}
\end{center}
\end{table*}
 
\begin{figure}[htbp]
\centering
\subfloat[]{\includegraphics[width=0.45\textwidth]{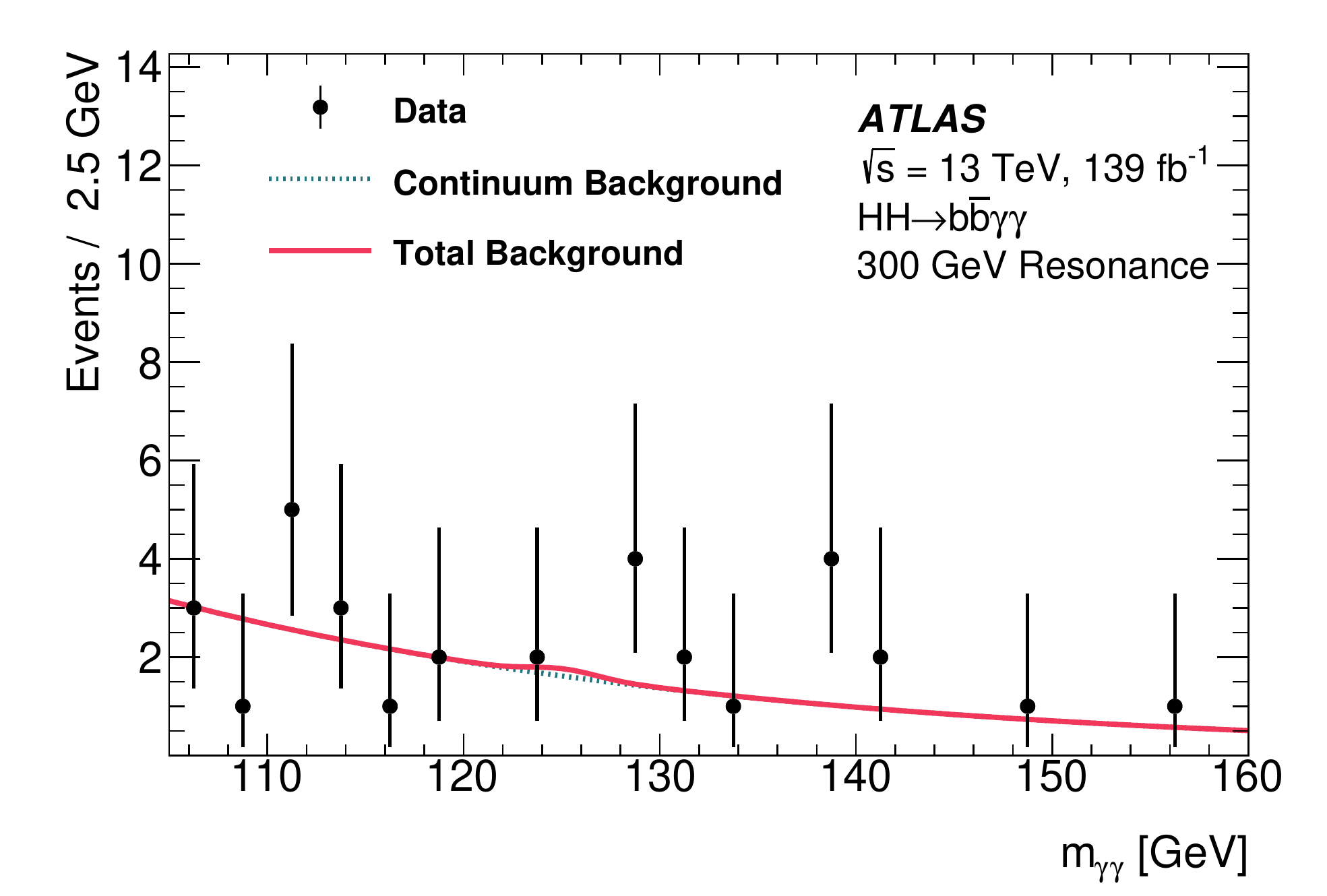}}
\subfloat[]{\includegraphics[width=0.45\textwidth]{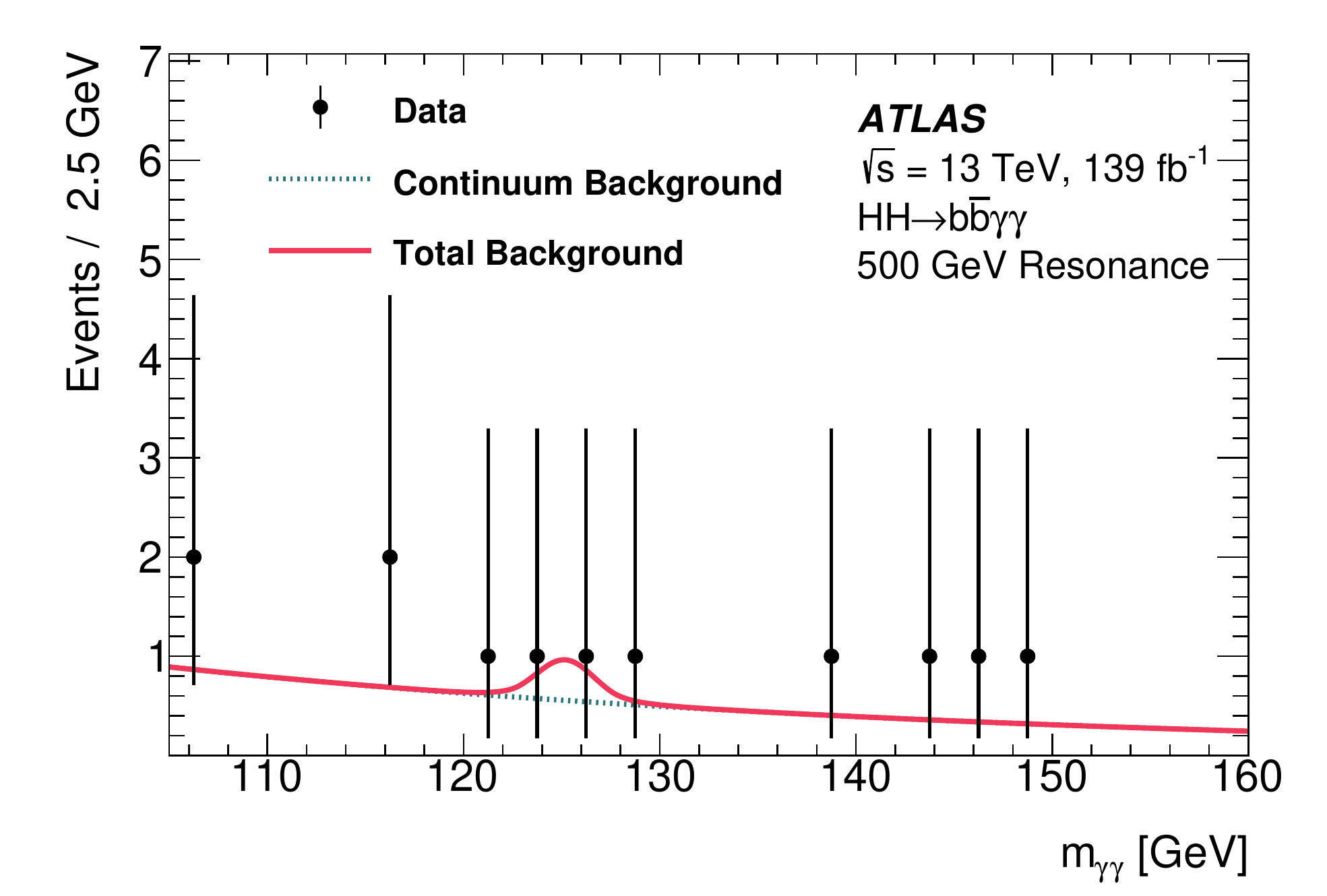}}
\caption{Data are compared with the background-only fit for the resonant search  for the (a) $m_{X}= 300~\GeV$ and (b) $m_{X}= 500~\GeV$ mass hypotheses. The continuum background, as well as the background from single Higgs boson production and from SM $HH$ production, is considered.  The data points in the Figure are the same as shown in Figure~\ref{fig:dataMC_myy_Res}.}
\label{fig:res_sideband_fit}
\end{figure}

\begin{figure}[htbp]
\centering
\includegraphics[width=0.65\textwidth]{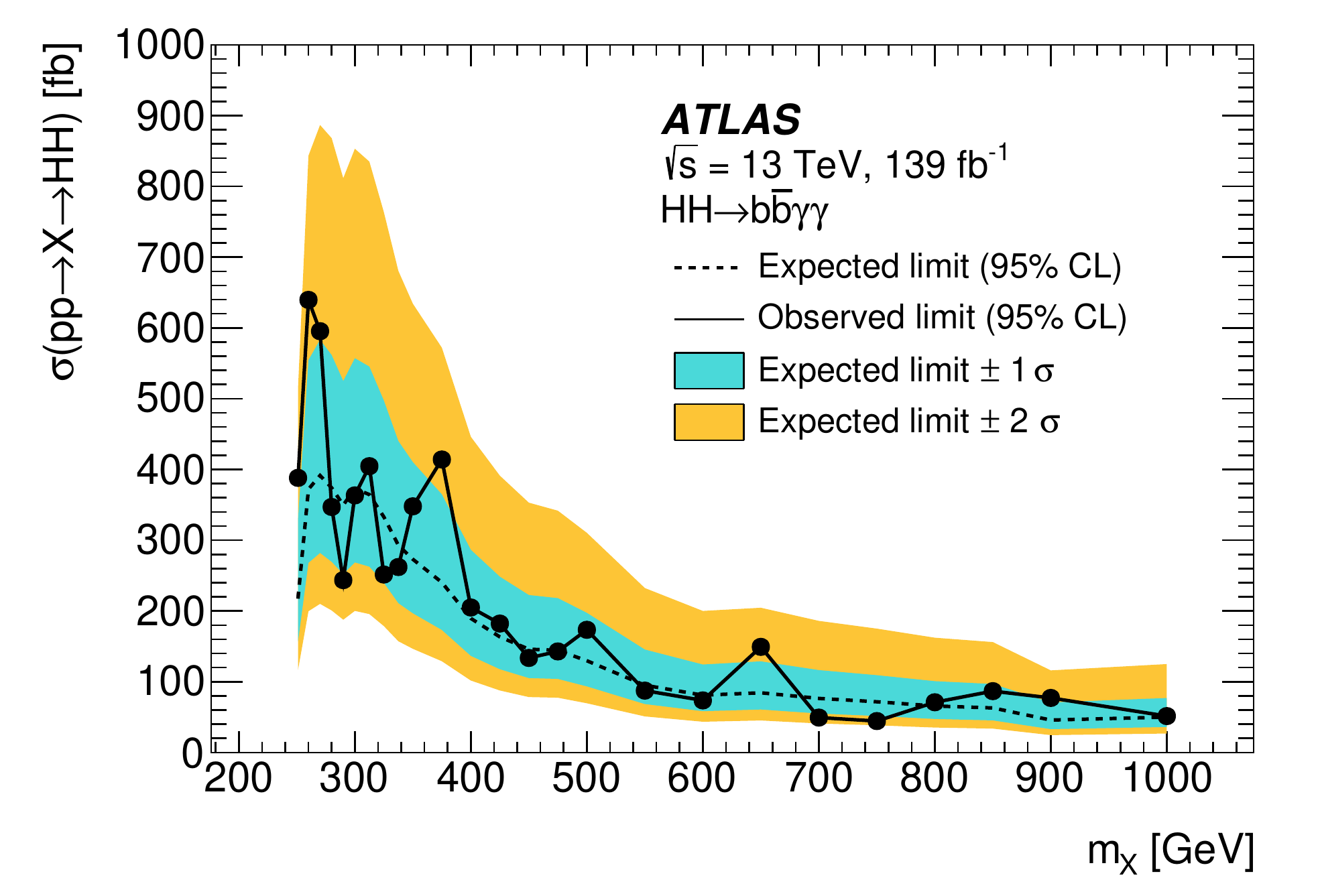}
\caption{Observed and expected limits at 95\% CL on the production cross section  of a narrow-width scalar resonance $X$ as a function of the mass $m_{X}$ of the hypothetical scalar particle. The black solid line represents the observed upper limits. The dashed line represents the expected upper limits. The $\pm 1\sigma$ and $\pm 2\sigma$  variations about the expected limit  due to statistical and systematic uncertainties are also shown.}
\label{fig:res_expected_limit}
\end{figure}

\FloatBarrier
\subsection{Impact of systematic uncertainties}
\label{sec:syst_impact}
The dominant systematic uncertainties are listed in Table~\ref{tab:syst} for both the nonresonant and resonant searches. The main uncertainties are related to the choice of functional form for the continuum background (spurious signal), to the parton showering model, and to the photon energy resolution.
 
\begin{table*}[!htb]
\footnotesize
\caption{Breakdown of the dominant systematic uncertainties.
The impact of the uncertainties
corresponds to the relative variation of the expected upper limit on the cross section when re-evaluating the profile likelihood ratio after fixing the nuisance parameter in question to its best-fit value, while all remaining nuisance parameters remain free to float.
The impact is shown in \%. Only systematic uncertainties with an impact of at least 0.2\% are shown. Uncertainties of the ``Norm.\ + Shape'' type affect both the normalization and the parameters of the functional form. The rest of the uncertainties affect only the yields. 
}
\begin{center}
\begin{tabular}{llcc}
\toprule
\toprule
&   & \multicolumn{2}{c}{Relative impact of the systematic uncertainties [\%]} \\
\midrule
Source &  Type & \multicolumn{1}{c}{Nonresonant analysis} &  \multicolumn{1}{c}{Resonant analysis}  \\
&       & $HH$    &   $m_{X}=300~\GeV$ \\
\midrule
Experimental \\
\midrule
Photon energy resolution & Norm.\ + Shape & 0.4 &  0.6 \\ 
Jet energy scale and resolution & Normalization & $<0.2$~~~~ & 0.3 \\
Flavor tagging & Normalization &  $<0.2$~~~~ & 0.2 \\
\midrule
Theoretical \\
\midrule
Factorization and renormalization scale & Normalization & 0.3 &  $<0.2$~~~~ \\
Parton showering model & Norm.\ + Shape & 0.6 & 2.6 \\
Heavy-flavor content   & Normalization & 0.3 & $<0.2$~~~~ \\
$\mathcal{B}$($H\rightarrow \gamma\gamma, b\bar{b}$) &  Normalization & 0.2 &  $<0.2$~~~~ \\
\midrule
Spurious signal & Normalization & 3.0 & 3.3\\
\bottomrule
\bottomrule
\end{tabular}
\label{tab:syst}
\end{center}
\end{table*}

\FloatBarrier
 
\section{Conclusions}
Searches for nonresonant and resonant Higgs boson pair production are performed in the \yybb{} final state using 139~fb$^{-1}$ of 13~\TeV{} $pp$ collision data collected with the ATLAS detector at the LHC. No significant excess above the Standard Model background expectation is observed. A 95\% CL upper limit of 130~fb is set on the $pp \rightarrow HH$ nonresonant production cross section, where the expected limit is 180~fb. The observed (expected) limit corresponds to 4.2 (5.7) times the cross section predicted by the Standard Model. Constraints on the Higgs boson self-coupling are also derived and limits of $-1.5< \kappa_{\lambda}<6.7$ are obtained, where  $-2.4< \kappa_{\lambda}<7.7$ is expected. The  expected constraints on the $HH$ nonresonant production cross section and on \klambda are obtained with a background hypothesis excluding $pp \rightarrow HH$ production.
For resonant production of a scalar particle $X\to\HH\to\yybb$, upper limits on the production cross section are obtained for the narrow-width hypothesis as a function of $m_{X}$. The observed (expected) upper limits are in the range  640--44~fb (391--46~fb) for $251 \leq m_{X} \leq 1000$~\GeV. Compared to the previous ATLAS result based on 36~fb$^{-1}$ of 13~\TeV{} $pp$ collisions, the present analysis uses a data set about four times larger, incorporates a categorization based on $\myybb^{*}$ and multivariate event selections, and expands the analyzed mass range of the resonance search to lower values. The results improve upon the previous ATLAS limits on the \hhyybb{} production cross section by up to a factor of five, and the allowed $\kappa_{\lambda}$ range shrinks by about a factor of two. 
For the resonant search, the expected limit on the cross section improves by a factor of two to three depending on the $m_X$ value. Of those improvements,  a factor of two arises from the increase in integrated luminosity, while the additional improvement can be attributed to the use of multivariate techniques, more precise object reconstruction and calibration and, for the nonresonant search, the categorization based on $\myybb^{*}$. The present analysis also sets constraints that are tighter than those from a combination of ATLAS searches for $HH$ production in up to 36~fb$^{-1}$ of 13~\TeV{} data.

% End of text imported from the .//sections/Results.tex input file
 
% The next lines are included from the .//sections/acknowledgements.tex input file

\section*{Acknowledgments}
% The next lines are included from the .//acknowledgements/Acknowledgements.tex input file

We thank CERN for the very successful operation of the LHC, as well as the
support staff from our institutions without whom ATLAS could not be
operated efficiently.
 
We acknowledge the support of
ANPCyT, Argentina;
YerPhI, Armenia;
ARC, Australia;
BMWFW and FWF, Austria;
ANAS, Azerbaijan;
SSTC, Belarus;
CNPq and FAPESP, Brazil;
NSERC, NRC and CFI, Canada;
CERN;
ANID, Chile;
CAS, MOST and NSFC, China;
Minciencias, Colombia;
MEYS CR, Czech Republic;
DNRF and DNSRC, Denmark;
IN2P3-CNRS and CEA-DRF/IRFU, France;
SRNSFG, Georgia;
BMBF, HGF and MPG, Germany;
GSRI, Greece;
RGC and Hong Kong SAR, China;
ISF and Benoziyo Center, Israel;
INFN, Italy;
MEXT and JSPS, Japan;
CNRST, Morocco;
NWO, Netherlands;
RCN, Norway;
MEiN, Poland;
FCT, Portugal;
MNE/IFA, Romania;
JINR;
MES of Russia and NRC KI, Russian Federation;
MESTD, Serbia;
MSSR, Slovakia;
ARRS and MIZ\v{S}, Slovenia;
DSI/NRF, South Africa;
MICINN, Spain;
SRC and Wallenberg Foundation, Sweden;
SERI, SNSF and Cantons of Bern and Geneva, Switzerland;
MOST, Taiwan;
TAEK, Turkey;
STFC, United Kingdom;
DOE and NSF, United States of America.
In addition, individual groups and members have received support from
BCKDF, CANARIE, Compute Canada and CRC, Canada;
COST, ERC, ERDF, Horizon 2020 and Marie Sk{\l}odowska-Curie Actions, European Union;
Investissements d'Avenir Labex, Investissements d'Avenir Idex and ANR, France;
DFG and AvH Foundation, Germany;
Herakleitos, Thales and Aristeia programmes co-financed by EU-ESF and the Greek NSRF, Greece;
BSF-NSF and GIF, Israel;
Norwegian Financial Mechanism 2014-2021, Norway;
NCN and NAWA, Poland;
La Caixa Banking Foundation, CERCA Programme Generalitat de Catalunya and PROMETEO and GenT Programmes Generalitat Valenciana, Spain;
G\"{o}ran Gustafssons Stiftelse, Sweden;
The Royal Society and Leverhulme Trust, United Kingdom.
 
The crucial computing support from all WLCG partners is acknowledged gratefully, in particular from CERN, the ATLAS Tier-1 facilities at TRIUMF (Canada), NDGF (Denmark, Norway, Sweden), CC-IN2P3 (France), KIT/GridKA (Germany), INFN-CNAF (Italy), NL-T1 (Netherlands), PIC (Spain), ASGC (Taiwan), RAL (UK) and BNL (USA), the Tier-2 facilities worldwide and large non-WLCG resource providers. Major contributors of computing resources are listed in Ref.~\cite{ATL-SOFT-PUB-2021-003}.
 
% End of text imported from the .//acknowledgements/Acknowledgements.tex input file
% End of text imported from the .//sections/acknowledgements.tex input file

\clearpage

\printbibliography

\clearpage
% ATLAS Collaboration author list
% Reference date of HDBS-2018-34 is 2021-03-11
% Author list last updated on date 12-SEP-22
% Data extracted on 12-Sep-2022 for paper reference HDBS-2018-34
% at 11:58am
 
\begin{flushleft}
\hypersetup{urlcolor=black}
{\Large The ATLAS Collaboration}

\bigskip

\AtlasOrcid[0000-0002-6665-4934]{G.~Aad}$^\textrm{\scriptsize 99}$,    
\AtlasOrcid[0000-0002-5888-2734]{B.~Abbott}$^\textrm{\scriptsize 124}$,    
\AtlasOrcid[0000-0002-7248-3203]{D.C.~Abbott}$^\textrm{\scriptsize 100}$,    
\AtlasOrcid[0000-0002-2788-3822]{A.~Abed~Abud}$^\textrm{\scriptsize 34}$,    
\AtlasOrcid[0000-0002-1002-1652]{K.~Abeling}$^\textrm{\scriptsize 51}$,    
\AtlasOrcid[0000-0002-2987-4006]{D.K.~Abhayasinghe}$^\textrm{\scriptsize 91}$,    
\AtlasOrcid[0000-0002-8496-9294]{S.H.~Abidi}$^\textrm{\scriptsize 27}$,    
\AtlasOrcid[0000-0001-5329-6640]{H.~Abramowicz}$^\textrm{\scriptsize 157}$,    
\AtlasOrcid[0000-0002-1599-2896]{H.~Abreu}$^\textrm{\scriptsize 156}$,    
\AtlasOrcid[0000-0003-0403-3697]{Y.~Abulaiti}$^\textrm{\scriptsize 5}$
% ...
\bigskip
\\

% ...
$^{5}$High Energy Physics Division, Argonne National Laboratory, Argonne IL; United States of America.\\
% ...
$^{27}$Physics Department, Brookhaven National Laboratory, Upton NY; United States of America.\\
% ...
$^{34}$CERN, Geneva; Switzerland.\\
% ...
$^{51}$II. Physikalisches Institut, Georg-August-Universit\"{a}t G\"ottingen, G\"ottingen; Germany.\\
% ...
$^{91}$Department of Physics, Royal Holloway University of London, Egham; United Kingdom.\\
% ...
$^{99}$CPPM, Aix-Marseille Universit\'e, CNRS/IN2P3, Marseille; France.\\
$^{100}$Department of Physics, University of Massachusetts, Amherst MA; United States of America.\\
% ...
$^{124}$Homer L. Dodge Department of Physics and Astronomy, University of Oklahoma, Norman OK; United States of America.\\
% ...
$^{156}$Department of Physics, Technion, Israel Institute of Technology, Haifa; Israel.\\
$^{157}$Raymond and Beverly Sackler School of Physics and Astronomy, Tel Aviv University, Tel Aviv; Israel.\\
% ...

\end{flushleft}

% Created with Glance <Atlas.Glance@cern.ch>

\end{document}